\documentclass[aps,prd,onecolumn,preprintnumbers,groupedaddress,showpacs,nofootinbib,amssymb]{revtex4-2}
\usepackage{graphicx,color}
\usepackage{amsmath}
\usepackage{amssymb}
\usepackage{amsfonts}
\usepackage{bm}
\usepackage{cancel}
\usepackage{slashed}
\newcommand{\e}{\mathrm{e}}

\usepackage{graphicx,color}
\usepackage{amsmath}
\usepackage{amssymb}
\usepackage{amsfonts}
\usepackage{bm}
\usepackage{cancel}
\usepackage{slashed}
\usepackage{hyperref}

\usepackage{tikz}
\usetikzlibrary{arrows.meta}
\usetikzlibrary{decorations.markings}
\usetikzlibrary{calc}

\usepackage{pgfplots}
\pgfplotsset{compat=1.18}

\allowdisplaybreaks[4]
\begin{document}

\preprint{KEK-TH-2890, KEK-Cosmo-0438}

\title{Dark Energy Based on Modified Gravity: Where the Future Lies?}

\author{Shin'ichi~Nojiri$^{1,2}$}
\email{nojiri@nagoya-u.jp}

\author{Sergei~D.~Odintsov$^{3,4}$}
\email{odintsov@ice.csic.es}

\author{V.K. Oikonomou$^{5,6}$}
\email{voikonomou@gapps.auth.gr}

\author{Oem Trivedi$^{7}$}
\email{oem.trivedi@vanderbilt.edu}

\affiliation{$^{1)}$ KEK Theory Center, Institute of Particle and Nuclear Studies, High Energy Accelerator Research Organization (KEK), Oho 1-1, Tsukuba, Ibaraki 305-0801, Japan}

\affiliation{$^{2)}$ Kobayashi-Maskawa Institute for the Origin of Particles and the Universe, Nagoya University, Nagoya 464-8602, Japan}

\affiliation{$^{3)}$ Institute of Space Sciences,
(ICE/CSIC-IEEC), Carrer de Can Magrans s/n, 08193 Barcelona, Spain}

\affiliation{$^{4)}$ Instituci\'o Catalana de Recerca i Estudis Avan\c{c}ats (ICREA),
Passeig Luis Companys, 23, 08010 Barcelona, Spain}

\affiliation{$^{5)}$Department of Physics, Aristotle University of
Thessaloniki, Thessaloniki 54124, Greece}

\affiliation{$^{6)}$Center for Theoretical Physics, Khazar
University, 41 Mehseti Str., Baku, AZ-1096, Azerbaijan}

\affiliation{$^{7)}$Department of Physics and Astronomy, Vanderbilt University, Nashville, TN 37235, USA}

\begin{abstract}
Based on current observational data, there is a strong possibility that dark energy was in a phantom regime in the near past, and a transition from a phantom to a quintessential regime may have occurred. 
Since then, many scenarios compatible with the DESI data have been proposed. 
In particular, there exist several realistic models in the framework of $F(R)$ gravity. 
In this paper, we show that such realistic models of $F(R)$ gravity may produce an anti-gravity or ghost graviton in the future. 
The problem could be avoided by further changing the parameters and fine-tuning the initial conditions, but we propose to introduce a potential barrier of the scalar field for the small scalar curvature region, which prevents the Universe from evolving into the anti-gravity or having a ghost graviton region, without the fine-tuning of the initial conditions. We also discuss how this regime can be tested using peculiar velocity and kSZ effect probes.
\end{abstract}

\maketitle

\section{Introduction\label{Sec1}}

In \cite{Nojiri:2025low}, the possibilities of a dynamically oscillating and phantom-crossing dark energy era have been considered in the framework of $F(R)$ gravity (for phantom-crossing dark energy in scalar-tensor and EGB gravity, see also \cite{Nojiri:2026uvn, Nojiri:2025uew}). 
Recently, the Dark Energy Spectroscopic Instrument (DESI) \cite{DESI:2024mwx, DESI:2025zgx} indicated that dark energy may have been in a phantom regime in the recent past. 
The DESI data might indicate the existence of a possible transition from a phantom to a quintessential regime in the framework of general relativity. 
Many proposals might produce results compatible with the DESI data, for example~\cite{Odintsov:2024woi, Dai:2020rfo, He:2020zns, Nakai:2020oit, DiValentino:2020naf, Agrawal:2019dlm, Ye:2020btb, Vagnozzi:2021tjv, Desmond:2019ygn, Hogas:2023pjz, OColgain:2018czj, Vagnozzi:2019ezj, Krishnan:2020obg, Colgain:2019joh, Vagnozzi:2021gjh, Lee:2022cyh, Krishnan:2021dyb, Ye:2021iwa, Ye:2022afu, Verde:2019ivm, Menci:2024rbq, Adil:2023ara, Reeves:2022aoi, Ferlito:2022mok, Vagnozzi:2021quy, DiValentino:2020evt, Sabogal:2024yha, Giare:2024smz, CosmoVerseNetwork:2025alb, Odintsov:2025kyw, vanderWesthuizen:2025iam, Paliathanasis:2025dcr, Odintsov:2025jfq, Kessler:2025kju, Nojiri:2025low}. 
There are recent researches using the $F(R)$ gravity~\cite{Nojiri:2017ncd, Capozziello:2011et, Faraoni:2010pgm, Nojiri:2006ri, Nojiri:2010wj, Nojiri:2003ft, Capozziello:2005ku, Hwang:2001pu, Song:2006ej, Faulkner:2006ub, Olmo:2006eh, Sawicki:2007tf, Faraoni:2007yn, Carloni:2007yv, Nojiri:2007as, Deruelle:2007pt, Appleby:2008tv, Dunsby:2010wg, Oikonomou:2025qub,Trivedi:2023zlf}, which might give models compatible with the DESI data~\cite{Odintsov:2025jfq, Nojiri:2025low}. 
In \cite{Kessler:2025kju}, it was also suggested that dark energy might have an oscillating equation-of-state (EoS) parameter. 
The possibility of oscillating dark energy was discussed in the framework of $F(R)$ gravity in \cite{Nojiri:2025low}.

A cosmological perfect fluid is called phantom \cite{Caldwell:1999ew, Caldwell:2003vq} when its EoS parameter $w$, which is the ratio of the pressure and the energy density of the fluid, is less than $-1$. 
Originally, the phantom crossing was defined as the transition from $w>-1$ to $w<-1$~\cite{Nojiri:2005pu, Vikman:2004dc, Hu:2004kh, Bamba:2008hq, Bamba:2012ky, Saitou:2012xw}, although the phantom crossing suggested by DESI is the transition from $w>-1$ to $w<-1$. 
There have been strong activities in the research of the phantom~\cite{Carroll:2003st, Cline:2003gs, Nojiri:2005sx, Feng:2004ad, Nojiri:2005sr, Nojiri:2005vv, Nojiri:2005jg, Kunz:2006wc, Bamba:2008hq, Nojiri:2003vn, Nojiri:2003jn, Nojiri:2003ag, Elizalde:2004mq, Nojiri:2005pu, Capozziello:2005tf, Elizalde:2005ju, Nojiri:2005am, Nojiri:2005am, Briscese:2006xu, Elizalde:2008yf, Bamba:2010bm, Astashenok:2012tv, Nojiri:2013ru, Nojiri:2015fia, Cardenas:2020grl}, see for \cite{Cai:2009zp} for the review of hantom crossing.

The $F(R)$ gravity was proposed to explain inflation~\cite{Starobinsky:1980te}. 
After that, models of dark energy have been proposed~\cite{Capozziello:2002rd, Nojiri:2003ft}. 
The models unifying inflation and dark energy have also been actively investigated \cite{Nojiri:2003ft, Nojiri:2007as, Nojiri:2007cq, Cognola:2007zu, Appleby:2009uf, Elizalde:2010ts, Nojiri:2009kx, Odintsov:2017hbk, Nojiri:2019fft, Odintsov:2025jfq} (see \cite{Nojiri:2010wj} for a review). 
Then $F(R)$ gravity could describe the evolution from the early Universe to the present one. 
Then a natural question is what could happen in the future. 
In principle, there are several possibilities:
\begin{itemize}
\item If the expanding Universe begins to contract,
\begin{itemize}
\item The Universe goes to the Big Crunch.
\item The Universe continues to contract forever. 
\end{itemize}
\item If the Universe continues to expand,
\begin{itemize}
\item The Universe goes to the asymptotic de~Sitter spacetime, including the pseudo-rip~\cite{Frampton:2011aa}, where $w$ increases to $-1$. 
\item The Universe continues to expand acceleratedly, but does not go to the asymptotic de~Sitter spacetime. 
\item The Universe goes to a little rip~\cite{Frampton:2011sp, Brevik:2011mm, Nojiri:2011kd, Frampton:2011rh}, where the Hubble rate goes to infinity in the infinite future. 
\item The Universe goes to a singularity in the finite future as classified in \cite{Nojiri:2005sx}. 
\end{itemize}
\end{itemize}
As we will see in this paper, the anti-gravity and ghost graviton appear in the finite future even in realistic models of $F(R)$ gravity~\cite{Odintsov:2025jfq, Odintsov:2024lid, Odintsov:2026fyc} and therefore the models become physically inconsistent. 
In these models, by fine-tuning the initial conditions, the Universe goes to the stable de~Sitter spacetime, and the models may solve the anti-gravity or ghost graviton problem. 
In order to avoid this problem, we propose to modify the models so that the potential barrier could appear in the future, but before the Universe goes into the anti-gravity or ghost graviton region. 
This is a rather general prescription. 

We may consider models that realize the evolution of the Universe such that after the (inverse) phantom crossing, it goes to the asymptotic de~Sitter spacetime or the decelerating Universe. 
As is well known, $F(R)$ gravity is rewritten in the form of a scalar-tensor theory in the Einstein frame. 
Although the cosmological time in the original Jordan frame differs from that in the Einstein frame, the Universe's evolution is governed by the scalar field potential in the Einstein frame. 
The scalar field rolls down to the lower region of the potential, and if there is a minimum in the potential, the minimum becomes a stable point for the scalar field; if the potential minimum is positive, the Universe goes to an asymptotic de~Sitter spacetime. 

In the next section, mainly based on \cite{Nojiri:2003ft, Nojiri:2006ri, Nojiri:2010wj, Nojiri:2017ncd}, we recall the general properties of $F(R)$ gravity. 
We use these properties in the arguments that follow. 
Section~\ref{Sec3} is also a review of Ref.~\cite{Nojiri:2025low}, where we show that the phantom crossing is a rather general phenomenon in $F(R)$ gravity theories and can be realized by properly choosing the initial conditions. 
In Section \ref{Sec4}, by starting to review the scalar-tensor theory description of $F(R)$ gravity and by using the potential of the corresponding scalar field, we discuss the future evolution of
the Universe. 
In Section~\ref{Sec5}, we investigate how matter contributes to the evolution of the Universe, especially in the future. Here we also discuss the behavior of matter in the Jordan frame. 
In Section~\ref{Sec6}, we investigate the future evolution of the Universe in some realistic models proposed in \cite{Odintsov:2026fyc}. 
We show that the anti-gravity or ghost graviton often appears in the future for these models. 
The problem might be avoided by further changing the parameters and fine-tuning the initial conditions. 
This problem is solved in Section~\ref{Sec7} without fine-tuning of the initial conditions. 
In Section~\ref{sec8}, we discuss a non-trivial observational outlook attached to this scenario, wherein we shall discuss how one can use peculiar velocity and kSZ effect probes for diagnosing this scenario. 
The problem can be avoided in general cases by introducing the potential barrier of the scalar field, which prevents the Universe from evolving into the anti-gravity or ghost graviton region. 
The last section is devoted to the summary and discussion. 

\section{General Properties of $F(R)$ gravity\label{Sec2}}

In this section, we review the general properties of $F(R)$ gravity based on \cite{Nojiri:2003ft, Nojiri:2006ri, Nojiri:2010wj, Nojiri:2017ncd}.
In particular, we show how some of the exact solutions in $F(R)$ gravity appear.

In $F(R)$ gravity, the scalar curvature $R$ in the Einstein-Hilbert action\footnote{
We use the following convention for the curvatures:
\[
R=g^{\mu\nu}R_{\mu\nu} \, , \quad
R_{\mu\nu} = R^\lambda_{\ \mu\lambda\nu} \, , \quad
R^\lambda_{\ \mu\rho\nu} = -\Gamma^\lambda_{\mu\rho,\nu}
+ \Gamma^\lambda_{\mu\nu,\rho} -
\Gamma^\eta_{\mu\rho}\Gamma^\lambda_{\nu\eta}
+ \Gamma^\eta_{\mu\nu}\Gamma^\lambda_{\rho\eta} \, ,\quad
\Gamma^\eta_{\mu\lambda} = \frac{1}{2}g^{\eta\nu}\left(
g_{\mu\nu,\lambda} + g_{\lambda\nu,\mu} - g_{\mu\lambda,\nu}
\right)\, .
\]
}
is replaced by an appropriate function of the scalar curvature,
\begin{align}
\label{JGRG7}
S_{F(R)}= \int d^4 x \sqrt{-g} \left( \frac{F(R)}{2\kappa^2} + \mathcal{L}_\mathrm{matter} \right)\, .
\end{align}
The equation for modified gravity is given by,
\begin{align}
\label{JGRG13}
\frac{1}{2}g_{\mu\nu} F(R) - R_{\mu\nu} F'(R) - g_{\mu\nu} \Box F'(R) + \nabla_\mu \nabla_\nu F'(R) = - \frac{\kappa^2}{2}T_{\mathrm{matter}\, \mu\nu}\, .
\end{align}
This equation corresponds to the Einstein equation in Einstein's gravity.

We assume a spatially flat Friedmann-Lema\^{i}tre-Robertson-Walker (FLRW) Universe,
\begin{align}
\label{JGRG14}
ds^2 = - dt^2 + a(t)^2 \sum_{i=1,2,3} \left(dx^i\right)^2\, .
\end{align}
Even for modified gravity, one may define an effective EoS parameter as follows,
\begin{align}
\label{JGRG12}
w_\mathrm{eff}= - 1 - \frac{2\dot H}{3H^2} \, .
\end{align}
Here $H=\frac{\dot a}{a}$. 
Assuming a spatially flat FLRW Universe~\eqref{JGRG14}, Eq.~\eqref{JGRG13} gives the equations corresponding to the Friedmann equations as follows,
\begin{align}
\label{JGRG15}
0 =&\,  -\frac{F(R)}{2} + 3\left(H^2 + \dot H\right) F'(R)  - 18 \left( 4H^2 \dot H + H \ddot H\right) F''(R) + \kappa^2 \rho_\mathrm{matter}\, ,\\
\label{Cr4b}
0 =&\, \frac{F(R)}{2} - \left(\dot H + 3H^2\right)F'(R) + 6 \left( 8H^2 \dot H + 4 {\dot H}^2 + 6 H \ddot H + \dddot H\right) F''(R) + 36\left( 4H\dot H + \ddot H\right)^2 F'''(R) \nonumber \\
&\, + \kappa^2 p_\mathrm{matter}\, .
\end{align}
Here, the scalar curvature $R$ is given by $R=12H^2 + 6\dot H$. 
By using \eqref{JGRG15} and \eqref{Cr4b}, we obtain a conservation law of matter,
\begin{align}
\label{cnsv}
0 = \dot \rho_\mathrm{matter} + 3 H \left( \rho_\mathrm{matter} + p_\mathrm{matter} \right)\, .
\end{align}
We often forget Eq.~\eqref{Cr4b} because we can also obtain it from Eq.~\eqref{JGRG15} and the conservation law~\eqref{cnsv}.

Several solutions of \eqref{JGRG15} have been found. 
Without any matter, assuming that the Ricci tensor is covariantly constant, that is, $R_{\mu\nu}\propto g_{\mu\nu}$, Eq.~\eqref{JGRG13} reduces to the algebraic equation,
\begin{align}
\label{JGRG16}
0 = 2 F(R) - R F'(R)\, .
\end{align}
If Eq.~(\ref{JGRG16}) has a solution, the (anti-)de~Sitter and/or Schwarzschild- (anti-)de~Sitter space, or Kerr - (anti-)de~Sitter spacetime is an exact vacuum solution. 

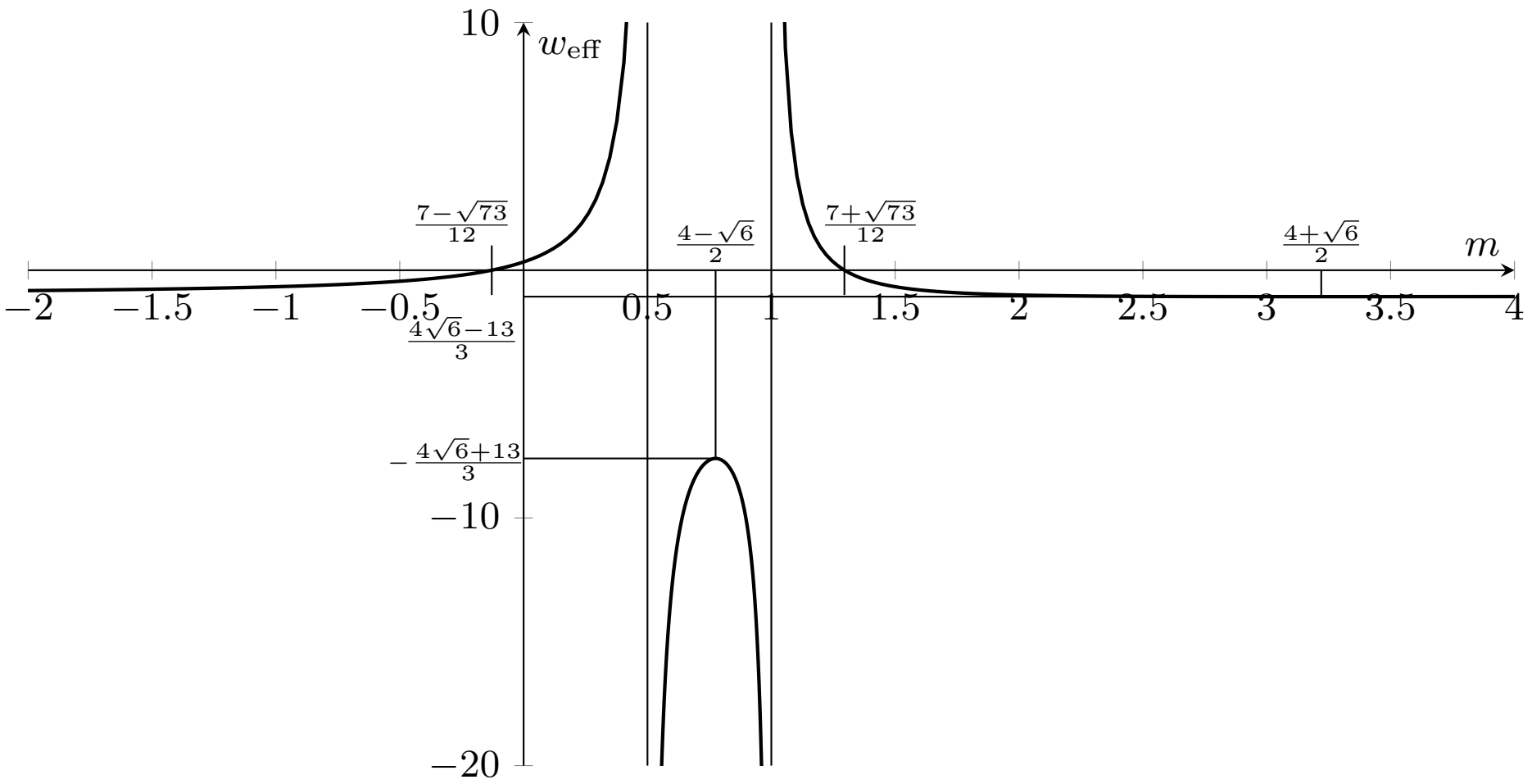
\begin{figure}[h]
\begin{center}

\begin{tikzpicture}
\begin{axis}[
x=2cm,
y=0.2cm,
axis lines=middle,
xmin=-2, xmax=4,
ymin=-20, ymax=10,
samples=300,
xlabel={{
$m$}},
ylabel={
{
$w_\mathrm{eff}$}},
]

\draw[-, thin] (-0.1286,-1) -- (-0.1286, 1);
\draw[-, thin] (1.295,-1) -- (1.295, 1);
\draw[-, thin] (1,-20) -- (1, 10);
\draw[-, thin] (0.5,-20) -- (0.5, 10);
\draw[-, thin] (0,-1.067) -- (3.22, -1.067);
\draw[-, thin] (3.22,0) -- (3.22, -1.067);
\draw[-, thin] (0,-7.599) -- (0.775, -7.599);
\draw[-, thin] (0.775,0) -- (0.775, -7.599);

\node at (-0.25, 2) {{\tiny $\frac{7 - \sqrt{73}}{12}$}};
\node at (1.4, 2) {{\tiny $\frac{7 + \sqrt{73}}{12}$}};

\node at (-0.25, -2.7) {{\tiny $\frac{4\sqrt{6} - 13}{3}$}};
\node at (3.22, 1.2) {{\tiny $\frac{4 + \sqrt{6}}{2}$}};

\node at (-0.27, -7.599) {{\tiny $-\frac{4\sqrt{6} + 13}{3}$}};
\node at (0.775, 1.2) {{\tiny $\frac{4 - \sqrt{6}}{2}$}};


\addplot[thick, domain=-8:0.49]
{ -(6*x*x - 7*x -1 )/(3*(x-1)*(2*x-1)) };

\addplot[thick, domain=0.505:0.99]
{ -(6*x*x - 7*x -1 )/(3*(x-1)*(2*x-1)) };

\addplot[thick, domain=1.01:8]
{ -(6*x*x - 7*x -1 )/(3*(x-1)*(2*x-1)) };

\end{axis}
\end{tikzpicture}

\end{center}

\caption{
The behavior of $w_\mathrm{eff}=-\frac{6m^2 - 7m - 1}{3(m-1)(2m -1)}$ in \eqref{JGRG18}.
$w_\mathrm{eff}$ vanishes at $m=\frac{7 + \sqrt{73}}{12} \fallingdotseq 1.295$ and $m= \frac{7 - \sqrt{73}}{12} \fallingdotseq - 0.1286$ and
$w_\mathrm{eff}$ diverges at $m=\frac{1}{2}$ and $m=1$.
$w_\mathrm{eff}$ has a local minimum $w_\mathrm{eff}=\frac{4\sqrt{6} - 13}{3}$ when $m=\frac{4 + \sqrt{6}}{2}$.
$w_\mathrm{eff}$ also has a local maximum $w_\mathrm{eff}= - \frac{4\sqrt{6} + 13}{3}$ when $m=\frac{4 - \sqrt{6}}{2}$.
We should also note $w_\mathrm{eff}\to -1$ when $\left| m \right| \to \infty$.
}
\label{weffm}
\end{figure}

We may assume that $F(R)$ behaves as,
\begin{align}
\label{Fpwr}
F(R) \sim  f_0 R^m\, .
\end{align}
Then Eq.~\eqref{JGRG15} gives
\begin{align}
\label{M7}
0 =&\, f_0 \left\{ - \frac{1}{2} \left(6\dot H + 12 H^2\right)^m + 3 m \left(\dot H + H^2\right)\left(6\dot H + 12 H^2\right)^{m - 1}
 - 3 m H \frac{d}{dt} \left\{\left(6\dot H + 12 H^2 \right)^{m -1}\right\}\right\} \nonumber \\
&\, + \kappa^2 \rho_0 a^{-3(1+w)}\, .
\end{align}
Eq.~\eqref{Cr4b} is irrelevant because it can be derived from \eqref{M7}.
When the contribution from the matter can be neglected ($\rho_0=0$), the following solution appears,
\begin{align}
\label{JGRG17}
H = \frac{-\frac{(m-1)(2m-1)}{m-2}}{t}\, ,
\end{align}
with a corresponding EoS parameter \eqref{JGRG12},
\begin{align}
\label{JGRG18}
w_\mathrm{eff}=-\frac{6m^2 - 7m - 1}{3(m-1)(2m -1)}
= - \frac{\left( m - \frac{7 + \sqrt{73}}{12} \right) \left( m - \frac{7 - \sqrt{73}}{12} \right)}{\left( m - 1 \right) \left(m - \frac{1}{2} \right)}
\fallingdotseq - \frac{\left( m - 1.295 \right) \left( m + 0.1286 \right)}{\left( m - 1 \right) \left( m - 0.5 \right)} \, .
\end{align}
The behaviour of $w_\mathrm{eff}$ as a function of $m$ is depicted in FIG.~\ref{weffm}. Eq.~\eqref{JGRG18} can be solved with respect to $m$, 
\begin{align}
\label{eqslv}
m= \frac{9 w_\mathrm{eff} + 7 \pm \sqrt{ 9 \left( w_\mathrm{eff} + \frac{13}{3} \right)^2 - 96}}{12 \left( w_\mathrm{eff} + 1 \right)} \, .\nonumber \\
\end{align}
Therefore $m$ is real when $w_\mathrm{eff} > \frac{4\sqrt{6} - 13}{3}= - 1.067\cdots $ or $w_\mathrm{eff} < - \frac{4\sqrt{6} + 13}{3}= - 7.599\cdots $ and $m$ is imaginary if $- \frac{4\sqrt{6}
+ 13}{3} < w_\mathrm{eff} < \frac{4\sqrt{6} - 13}{3}$. 
When $m$ is imaginary, we may consider the model, $F(R) \propto f_0 R^m + {f_0}^* R^{m^*}$. 
Here ``$*$'' denotes the complex conjugate. 
We should also note that the correspondence between $m$ and $w_\mathrm{eff}$ is two-to-one. 

When $w_\mathrm{eff} = \frac{4\sqrt{6} - 13}{3} = - 1.067\cdots $, $m$ is given by,
\begin{align}
\label{eqslvm}
m = \frac{4 + \sqrt{6}}{2}
= 3.22 \cdots \, .
\end{align}
On the other hand, when $w_\mathrm{eff} = - \frac{4\sqrt{6} + 13}{3}= - 7.599\cdots $, 
\begin{align}
\label{eqslvp}
m = \frac{4 - \sqrt{6}}{2}
= 0.775 \cdots \, .
\end{align}

When the matter contribution is included with a constant EoS parameter $w$, an exact solution of \eqref{M7} is given by
\begin{align}
\label{M8}
a=&\, a_0 t^{h_0} \, ,\quad h_0\equiv \frac{2m}{3(1+w)} \, ,\nonumber \\
a_0\equiv&\, \left[-\frac{3f_0h_0}{\kappa^2 \rho_0}\left(-6h_0 + 12 h_0^2\right)^{m-1}
\left\{\left(1-2m\right)\left(1-m\right) - (2-m)h_0\right\}\right]^{-\frac{1}{3(1+w)}}\, ,
\end{align}
and the effective EoS parameter \eqref{JGRG12} is,
\begin{align}
\label{JGRG20}
w_\mathrm{eff}= -1 + \frac{w+1}{m}\, .
\end{align}
Let us now introduce perfect fluid and scalar-tensor representations for the modified gravity under consideration. 
Eqs.~\eqref{JGRG15} and \eqref{Cr4b} show that the effective energy density $\rho_\mathrm{eff}$ and the effective pressure $p_\mathrm{eff}$ including the contribution from $f(R)$ are given by,
\begin{align}
\label{Cr4}
\rho_\mathrm{eff} =&\, \frac{1}{\kappa^2}\left(-\frac{1}{2}f(R) + 3\left(H^2 + \dot H\right) f'(R)  - 18 \left(4H^2 \dot H + H \ddot H\right)f''(R)\right) + \rho_\mathrm{matter}\, ,\\
\label{Cr4bb}
p_\mathrm{eff} =&\, \frac{1}{\kappa^2}\left(\frac{1}{2}f(R) - \left(3H^2 + \dot H \right)f'(R) + 6 \left(8H^2 \dot H + 4{\dot H}^2 + 6 H \ddot H + \dddot H \right)f''(R)
+ 36\left(4H\dot H + \ddot H\right)^2f'''(R) \right) \nonumber \\
&\, + p_\mathrm{matter}\, .
\end{align}
The equations \eqref{JGRG15}) and \eqref{Cr4b} can be rewritten as in Einstein's gravity case,
\begin{align}
\label{JGRG11B}
\rho_\mathrm{eff}=\frac{3}{\kappa^2}H^2 \, ,
\quad p_\mathrm{eff}= - \frac{1}{\kappa^2}\left(3H^2 + 2\dot H\right)\, .
\end{align}

\section{Phantom Crossing of general $F(R)$ gravity\label{Sec3}}

We now briefly review the mechanism of the phantom crossing. 
The phantom crossing, as in the DESI observations~\cite{DESI:2024mwx, DESI:2025zgx}, can occur in general $F(R)$ gravity if the initial conditions are chosen properly. 
This section is based on the arguments in \cite{Nojiri:2025low}. 

Let $w$ be the EoS parameter of the dark energy, which is the ratio of the pressure of the dark energy to the energy density of the dark energy. 
If we believe the recent DESI observations~\cite{DESI:2024mwx, DESI:2025zgx},  a transition from $w<-1$ to $w>-1$ might have happened.

The effective energy density $\rho_\mathrm{eff}$ in \eqref{Cr4} and the effective pressure $p_\mathrm{eff}$ in \eqref{Cr4bb} tell that the effective energy density $\rho_{f(R)}$ and the effective pressure $p_{f(R)}$ coming from the $f(R)$ part are given by, 
\begin{align}
\label{Cr4F}
\rho_{f(R)} =&\, \frac{1}{\kappa^2}\left(-\frac{1}{2}f(R) + 3\left(H^2 + \dot H\right) f'(R) - 18 \left(4H^2 \dot H + H \ddot H\right)f''(R)\right) \, ,\\
\label{Cr4bbF}
p_{f(R)} =&\, \frac{1}{\kappa^2}\left(\frac{1}{2}f(R) - \left(3H^2 + \dot H \right)f'(R) + 6 \left(8H^2 \dot H + 4{\dot H}^2 + 6 H \ddot H + \dddot H \right)f''(R) \right. \nonumber \\
&\, \left. + 36\left(4H\dot H + \ddot H\right)^2f'''(R) \right) \, ,
\end{align}
which also satisfy the conservation law,
\begin{align}
\label{c}
0 = \dot\rho_{f(R)} + 3 H \left( \rho_{f(R)} + p_{f(R)} \right)\, .
\end{align}
If the standard phantom crossing (from $w>-1$ to $w<-1$) or the inverse phantom crossing (from $w<-1$ to $w>-1$) occurs, we have $\rho_{f(R)} = - p_{f(R)}$, and therefore $\dot\rho_{f(R)}$ must vanish. 
Because in the phantom Universe, we have $\rho_{f(R)}<-p_{f(R)}$, and in the non-phantom Universe, $\rho_{f(R)}>p_{f(R)}$, the crossing from the non-phantom Universe to the phantom one occurs if $\ddot\rho_{f(R)}>0$ and an inverse crossing occurs if $\ddot\rho_{f(R)}<0$.

We rewrite Eq.~\eqref{JGRG15} in the form of the standard Friedmann equation,
\begin{align}
\label{1F}
\frac{3}{\kappa^2} H^2 = \rho_{f(R)} + \rho_\mathrm{matter} \, ,
\end{align}
We consider dust, which gives the matter energy density $\rho_\mathrm{matter}$, i.e., baryonic matter and cold dark matter.
The energy density behaves as,
\begin{align}
\label{rhom}
\rho_\mathrm{matter} = \rho_\mathrm{matter(0)} a^{-3} = \rho_\mathrm{matter(0)} \e^{-3N} \, .
\end{align}
Here $\rho_\mathrm{matter(0)}$ is a constant and $a$ is the scale factor in \eqref{JGRG14}.
The scale factor $a$ is given in terms of the $e$-folding number $N$ as $a=\e^N$.

We now assume that the phantom crossing in the $f(R)$ sector has happened at $t=t_\mathrm{ph}$ in cosmic time. 
We may expand $N$ as follows,
\begin{align}
\label{phN}
N=\sum_{n=0} \frac{H_n}{\left( n + 1 \right)!} \left( t - t_\mathrm{ph} \right)^{n+1}\, .
\end{align}
Then the Hubble rate $H$ is given by,
\begin{align}
\label{phH}
H=\sum_{n=0} \frac{H_n}{n !} \left( t - t_\mathrm{ph} \right)^n\, ,
\end{align}
and the scalar curvature $R$ behaves as,
\begin{align}
\label{phR}
R =&\, 12 H^2 + 6\dot H
= 12 {H_0}^2 + 6 H_1 + \left( 24 H_0 H_1 + 6 H_2 \right) \left( t - t_\mathrm{ph} \right) \nonumber \\
&\, + \left( 12 H_0 H_2 + 12 {H_1}^2 + 3 H_3 \right) \left( t - t_\mathrm{ph} \right)^2
+ \mathcal{O} \left( \left( t - t_\mathrm{ph} \right)^3 \right)\, .
\end{align}
Then the effective energy density $\rho_{f(R)}$ of the $f(R)$ sector is also expanded as,
\begin{align}
\label{rhofR}
\rho_{f(R)} = \rho_{{f(R)}\, (0)} + \rho_{{f(R)}\, (2)} \left( t - t_\mathrm{ph} \right)^2 + \mathcal{O} \left( \left( t - t_\mathrm{ph} \right)^3 \right)\, .
\end{align}
If $\rho_{{f(R)}\, (2)}>0$, a crossing from the non-phantom
$\rho_{f(R)} > p_{f(R)}$ to the phantom $\rho_{f(R)} < p_{f(R)}$
occurs and if $\rho_{{f(R)}\, (2)}<0$, an inverse crossing occurs.

The equation \eqref{1F} corresponding to the first Friedmann
equation indicates,
\begin{align}
\label{1F0}
\frac{3}{\kappa^2} {H_0}^2 =&\, \rho_{{f(R)}\, (0)} + \rho_\mathrm{matter(0)} \, , \\
\label{1F1}
\frac{6}{\kappa^2} H_0 H_1 =&\, - 3 \rho_\mathrm{matter(0)} H_0 \, , \\
\label{1F2}
\frac{3}{\kappa^2} \left( H_0 H_2 + 2 {H_1}^2 \right) =&\, \rho_{{f(R)}\, (2)} + \rho_\mathrm{matter(0)} \left( \frac{9}{2} {H_0}^2 - \frac{3}{2} H_1 \right) \, .
\end{align}
On the other hand, Eq.~\eqref{Cr4F} tells,
\begin{align}
\label{Cr4t1}
\rho_{f(R)\, (0)} =&\, \frac{1}{\kappa^2}\Bigl(-\frac{1}{2}f\left( 12 {H_0}^2 + 6 H_1 \right)
+ 3\left( {H_0}^2 + H_1 \right) f'\left( 12 {H_0}^2 + 6 H_1 \right) \nonumber \\
&\, - 18 \left(4{H_0}^2 H_1 + H_0 H_2 \right)f''\left( 12 {H_0}^2 + 6 H_1 \right) \Bigr) \, ,\\
\label{rhofR1}
0 =&\, \frac{1}{\kappa^2}\left\{
 - 6 H_0 H_1 f'\left( 12 {H_0}^2 + 6 H_1 \right) \right. \nonumber \\
&\, + 18 \left( 4 {H_0}^3 H_1 - 3 {H_0}^2 H_2 - 4 H_0 {H_1}^2 - H_0 H_3 \right) f''\left( 12 {H_0}^2 + 6 H_1 \right) \nonumber \\
&\, \left.  - 108 H_0 \left(4 H_0 H_1 + H_2 \right)^2 f'''\left( 12 {H_0}^2 + 6 H_1 \right) \right\} \, ,\\
\label{rhofR2}
\rho_{f(R)\, (2)} =&\, \frac{1}{\kappa^2}\left\{ - 9 H_0 H_2 f'\left( 12 {H_0}^2 + 6 H_1 \right) \right. \nonumber \\
&\, + \left( - 36 {H_0}^2 H_2 + 108 H_0 H_1 H_2 - 144 {H_1}^3 + 9 {H_2}~2 \right) f''\left( 12 {H_0}^2 + 6 H_1 \right) \nonumber \\
&\, + \left( 24 H_0 H_1 + 6 H_2 \right) \left( 36 {H_0}^3 H_1 - 63 {H_0}^2 H_2 - 108 H_0 {H_1}^2 - 18 H_0 H_3 - 9 H_1 H_2 \right) \nonumber \\
&\, \times f'''\left( 12 {H_0}^2 + 6 H_1 \right) \nonumber \\
&\, \left.  - 324 H_0 \left( 4 H_0 H_1 + H_2 \right)^3 f''''\left( 12 {H_0}^2 + 6 H_1 \right) \right\} \, .
\end{align}
Eq.~\eqref{1F1} tells
\begin{align}
\label{H1} H_1 = - \frac{\kappa^2}{2} \rho_\mathrm{matter(0)} \, .
\end{align}
Then we find that $H_1$ is always negative.
By combining \eqref{1F0}, \eqref{Cr4t1} and \eqref{H1}, we solve these equations with respect to $H_2$, as follows,
\begin{align}
\label{H2}
H_2 =&\, - 4 H_0 H_1 - \frac{f\left( 12 {H_0}^2 + 6 H_1 \right)}{36 H_0 f''\left( 12 {H_0}^2 + 6 H_1 \right)}
+ \frac{\left( {H_0}^2 + H_1 \right) f'\left( 12 {H_0}^2 + 6 H_1 \right)}{6H_0 f''\left( 12 {H_0}^2 + 6 H_1 \right)} \nonumber \\
&\, - \frac{3{H_0}^2 + H_1}{18 H_0 f''\left( 12 {H_0}^2 + 6 H_1 \right)} \, .
\end{align}
Eq.~\ref{1F2} with \eqref{H1} and \eqref{H2} also tell,
\begin{align}
\label{rho2}
\rho_{{f(R)}\, (2)} =&\, \frac{3}{\kappa^2} \left( H_0 H_2 + 2 {H_1}^2 \right)
+ \frac{2}{\kappa^2} H_1 \left( \frac{9}{2} {H_0}^2 - \frac{3}{2} H_1 \right) \nonumber \\
=&\, \frac{3}{\kappa^2} \left( - {H_0}^2 H_1 + {H_1}^2 \right)
 - \frac{f\left( 12 {H_0}^2 + 6 H_1 \right)}{12 \kappa^2 f''\left( 12 {H_0}^2 + 6 H_1 \right)} \nonumber \\
&\, + \frac{\left( {H_0}^2 + H_1 \right) f'\left( 12 {H_0}^2 + 6 H_1 \right)}{2 \kappa^2 f''\left( 12 {H_0}^2 + 6 H_1 \right)}
 - \frac{3{H_0}^2 + H_1}{6 \kappa^2 f''\left( 12 {H_0}^2 + 6 H_1 \right)} \, .
\end{align}
Therefore, if a crossing from the non-phantom to the phantom occurs, we find $\rho_{{f(R)}\, (2)}>0$, or if an inverse crossing occurs, we find $\rho_{{f(R)}\, (2)}<0$. 
Then we can find that the phantom crossing or the inverse phantom crossing occurs in a wide class of $F(R)$ gravity models, although it depends on the initial condition determining $H_0$.

We consider $R^2$ gravity as an example,
\begin{align}
\label{R2}
f(R)=\alpha R^2 \, .
\end{align}
Here $\alpha$ is a constant. 
As is well known, in $F(R)$ gravity in model (\ref{R2}), the square of the mass of the propagating scalar mode is $\frac{1}{\alpha}$, which implies that in order to avoid a tachyon, $\alpha$ must be positive.

For the model \eqref{R2}, Eq.~\eqref{rho2} has the following form,
\begin{align}
\label{rho2B0}
\rho_{{f(R)}\, (2)}=&\, \frac{1}{2\kappa^2} \left( 9 {H_1}^2 - \frac{3{H_0}^2 + H_1}{6 \alpha} \right) \, .
\end{align}
Hence the standard phantom crossing occurs, the condition is given
by,
\begin{align}
\label{sphcndtn} {H_0}^2 < 18 \alpha {H_1}^2 - \frac{1}{3} H_1 = \frac{9 \alpha \kappa^4}{2} {\rho_\mathrm{matter(0)}}^2 + \frac{\kappa^2}{6} \rho_\mathrm{matter(0)} \, .
\end{align}
Here, we used Eq. \eqref{H1}. On the other hand, the condition for
the inverse phantom crossing is given by,
\begin{align}
\label{iphcndtn}
{H_0}^2 > 18 \alpha {H_1}^2 - \frac{1}{3} H_1 = \frac{9 \alpha \kappa^4}{2} {\rho_\mathrm{matter(0)}}^2
+ \frac{\kappa^2}{6} \rho_\mathrm{matter(0)} \, .
\end{align}
This tells us that the crossing depends on the initial condition to determine $H_0$ and $\rho_\mathrm{matter(0)}$.

\section{Scalar-Tensor Theory Description of $F(R)$ gravity and Future evolution of Universe\label{Sec4}}

We would like to investigate the evolution of the Universe after the (inverse) phantom crossing. Because $F(R)$ gravity can be rewritten as a scalar-tensor theory in the Einstein frame, the Universe's evolution is governed by the scalar field potential in the Einstein frame. 
Especially the stability of the de~Sitter spacetime, which is often an exact solution of $F(R)$ gravity, can be determined from the potential of the scalar field.

One can also rewrite $F(R)$ gravity in the scalar-tensor form. 
By introducing the auxiliary field $A$, the action \eqref{JGRG7} of the $F(R)$ gravity is rewritten in the following form, 
\begin{align}
\label{JGRG21}
S=\frac{1}{2\kappa^2}\int d^4 x \sqrt{-g} \left\{F'(A)\left(R-A\right) + F(A)\right\}\, .
\end{align}
By the variation of the action~\eqref{JGRG21} with respect to $A$, one obtains $A=R$.
Substituting $A=R$ into the action \eqref{JGRG21}, one can reproduce the action in \eqref{JGRG7}.
Furthermore, we rescale the metric in the following way (field-dependent Weyl rescaling),
\begin{align}
\label{JGRG22}
g_{\mu\nu}\to \e^\sigma g_{\mu\nu}\, ,\quad \sigma = -\ln F'(A)\, .
\end{align}
(or $g_{\mu\nu}^\mathrm{Jordan\ frame}= \e^\sigma g_{\mu\nu}^\mathrm{Einstein\ frame}$).
Thus, the Einstein frame action is obtained:
\begin{align}
\label{JGRG23}
S_\mathrm{E} =&\, \frac{1}{2\kappa^2}\int d^4 x \sqrt{-g} \left( R - \frac{3}{2}g^{\rho\sigma} \partial_\rho \sigma \partial_\sigma \sigma - V(\sigma)\right) \, ,\nonumber \\
V(\sigma) =&\, \e^\sigma g\left(\e^{-\sigma}\right) - \e^{2\sigma} f\left(g\left(\e^{-\sigma}\right)\right)
= \frac{A}{F'(A)}  - \frac{F(A)}{F'(A)^2}\, .
\end{align}
Here $g\left(\e^{-\sigma}\right)$ is given by solving the equation $\sigma =- \ln F'(A)$ as $A=g\left(\e^{-\sigma}\right)$. 
Due to the field-dependent Weyl rescaling \eqref{JGRG22}, a coupling of the scalar field $\sigma$ with usual matter arises. 
The mass of $\sigma$ is given by, 
\begin{align}
\label{JGRG24}
{m_\sigma}^2 \equiv \frac{3}{2}\frac{d^2 V(\sigma)}{d\sigma^2}
=\frac{3}{2}\left\{\frac{A}{F'(A)} - \frac{4F(A)}{\left(F'(A)\right)^2} + \frac{1}{F''(A)}\right\}\, .
\end{align}

Note that the anti-gravity regime appears when $F'(R)$ is negative, which follows from Eq.~\eqref{JGRG21}. 
Then, we need to require
\begin{align}
\label{FR1}
F'(R) > 0 \, .
\end{align}
We should note that anti-gravity means that the graviton is a ghost in quantum theory because the signature of the kinetic term of the graviton changes, as is clear from \eqref{JGRG21}. 
The existence of the ghost conflicts with the framework of the quantum theory because it generates negative-norm states. 
Therefore, the theory including ghosts cannot be physically consistent and must be excluded. In the case of \eqref{Fpwr}, $F(R) \sim  f_0 R^m$,
Eq.~\eqref{FR1} require, 
\begin{align}
\label{Fpwrd}
m f_0 R^{m-1} > 0\, .
\end{align}
In cosmology, we find $R\geq 0$ in many cases and therefore we may require
\begin{align}
\label{Fpwrd2}
m f_0 > 0 \, .
\end{align}
Eq.~\eqref{JGRG22} also tells us that in the anti-gravity region, the scalar field $\sigma$ becomes complex because for any real and positive number $C$, $\ln (-C)=\ln C \pm i\pi$. 
And at the transition point $F'(A)=0$, the scalar field diverges, which corresponds to the branch point and singular point of the logarithmic function. 
This situation is a little bit similar, although it is different, to the case investigated in \cite{Briscese:2006xu}, where the $F(R)$ gravity theory corresponding to the phantom scalar becomes complex.

We should also note that
\begin{align}
\label{FRV1}
\frac{dV(\sigma)}{dA} = \frac{F''(A)}{F'(A)^3} \left( - AF'(A) + 2F(A) \right)\, .
\end{align}
Therefore, if
\begin{align}
\label{FRV2}
0 = - AF'(A) + 2F(A) \, ,
\end{align}
the scalar field $\sigma$ is at a local maximum or local minimum of the potential and, therefore, $\sigma$ can be a constant. 
Note that the condition (\ref{FRV2}) is nothing but the condition~\eqref{JGRG16} for the existence of the de~Sitter solution. 
Since $\e^{-\sigma}=F'(A)$, we have $\frac{dA}{d\sigma} = - \frac{F'(A)}{F''(A)}$ and therefore Eq.~\eqref{FRV1} tells,
\begin{align}
\label{FRV1B}
\frac{dV(\sigma)}{d\sigma} = \frac{A}{F'(A)} - \frac{2F(A)}{F'(A)^2}
= - \frac{A}{F'(A)} + 2V(\sigma)
= - A\e^\sigma + 2V(\sigma) \, .
\end{align}
When the condition \eqref{FRV2} is satisfied, the mass~\eqref{JGRG24} can be rewritten as 
\begin{align}
\label{FRV3}
{m_\sigma}^2 = \frac{3}{2 F'(A)} \left( - A + \frac{F'(A)}{F''(A)} \right)\, .
\end{align}
Then, when the condition \eqref{FR1} for the non-existence of the anti-gravity is satisfied, the mass squared ${m_\sigma}^2$ is positive and, therefore, the scalar field is at a local minimum if 
\begin{align}
\label{FRV4}
 - A + \frac{F'(A)}{F''(A)} > 0\, .
\end{align}
On the other hand, if,
\begin{align}
\label{FRV5}
 - A + \frac{F'(A)}{F''(A)} < 0\, ,
\end{align}
the scalar field is at a local maximum of the potential, and the mass squared ${m_\sigma}^2$ is negative. 
The condition \eqref{FRV4} is simply the stability condition for de~Sitter space. 
Therefore, if \eqref{FRV4} is satisfied, the Universe may evolve to an asymptotic de~Sitter spacetime.

For the model \eqref{Fpwr}, the potential in \eqref{JGRG23} is given by, 
\begin{align}
\label{VFpwr}
V = \frac{1}{m f_0} \left( 1 -  \frac{1}{m}\right) A^{2-m} \, .
\end{align}
In order to avoid the anti-gravity or ghost graviton, we require Eq.~\eqref{Fpwrd2}. Then $V$ is a monotonically increasing function of $A$ if $m<0$ or $1<m<2$ and a decreasing function if $0<m<1$ or $m>2$. Since $A$ is identified with the scalar curvature $R$ ($A=R$), if the scalar curvature decreases over time, $V$ also decreases; therefore, $V$ must be an increasing function of $A$, i.e., $m<0$ or $1<m<2$. 
In the phantom phase, the curvature becomes larger in the future; that is, $V$ must decrease with $A$, so $m<1$ or $m>2$.

As a model, we may consider the following:  
\begin{align}
\label{model1}
F(R) = R + f_0^e R^{m_e} + f_0^l R^{m_l}\, .
\end{align}
Here $f_0^e$, $f_0^l$, $m^e$, and $m^l$ are constant parameters and satisfying $m^e f_0^e >0$ and $m^l f_0^l>0$ as in \eqref{Fpwrd2}. 
The first term in \eqref{model1} corresponds to the Einstein-Hilbert term. 
If we choose $m_e>1$ and $m_l<1$, at early times when $R$ is larger, the second term in \eqref{model1} could dominate, and when $R$ is smaller, as in the late time, the last term could dominate. 
If we choose $m_e \lesssim 2$, we find $w_\mathrm{eff} \gtrsim -1$, and we obtain an accelerating Universe although it is not a phantom Universe. 
If we choose $m_l = \frac{7-\sqrt{73}}{12} \sim - 0.1286\cdots$, we obtain $w_\mathrm{eff} = 0$, and we obtain a dust Universe at late time. 
Anyway, $V$ becomes an increasing function of $A=R$, and therefore the curvature becomes smaller as time passes, as expected. 
In the model~\eqref{model1}, it looks like there could be no phantom crossing, and we might consider the apparent phantom crossing. 
As we have shown in the last section, the (inverse) phantom crossing can occur, which depends on the initial conditions. 
Therefore, even in the model~\eqref{model1}, there could be the (inverse) phantom crossing.

Let the solution of Eq.~\eqref{FRV2} be $A=A_0>0$.
Then Eq.~\eqref{FRV2} can be rewritten as $0=\left.\left( \frac{F(A)}{A^2} \right)'\right|_{A=A_0}$, whose solution is
\begin{align}
\label{slFRV2}
F(A) = F^{(1)}(A) \left( A - A_0 \right)^2 + F^{(0)} A^2 \, .
\end{align}
Here $F^{(0)}$ is a constant and $F^{(1)}(A)$ is a smooth function of $A$.
Because by assuming \eqref{FR1}, the condition \eqref{FRV4} can be rewritten as
\begin{align}
\label{FRV4R}
\left. \left( \frac{F'(A)}{A} \right)' \right|_{A=A_0} <0\ \mbox{and}\ F''\left(A_0\right)>0 \, ,
\end{align}
and for \eqref{slFRV2}, we have,
\begin{align}
\label{slFRV3}
F'\left(A_0\right) = 2 F^{(0)} A_0 \, , \quad
F''\left(A_0\right) = 2 F^{(1)}\left(A_0\right) + 2 F^{(0)} \, ,
\end{align}
we find
\begin{align}
\label{slFRV4}
F^{(0)} > 0 \, , \quad F^{(1)}\left(A_0\right) + F^{(0)}>0 \, , \quad
 - A_0 \left( 2 F^{(1)}\left(A_0\right) + 2 F^{(0)} \right) + 2 F^{(0)} A_0>0 \ \mbox{or} \ F^{(1)}\left(A_0\right)<0 \, .
\end{align}
We now rewrite \eqref{slFRV2} as follows,
\begin{align}
\label{slFRV5}
F(A) = - F^{(0)} {A_0}^2 + 2 F^{(0)} A_0 A + \left( F^{(0)} + F^{(1)}(A) \right) \left( A - A_0 \right)^2 \, .
\end{align}
The first term could be identified with a cosmological constant, and the second term could be regarded as the Einstein-Hilbert term by redefining the gravitational constant $\kappa$. 
If we choose $F^{(0)} + F^{(1)}(A)$ as an increasing function, which may dominate at late time or small $A=R$, the conditions \eqref{FR1}, \eqref{FRV2}, and \eqref{FRV4}. 
For example, we may consider the following,
\begin{align}
\label{slFRV6}
F^{(0)} + F^{(1)}(A) = F^{(2)} \tanh \frac{A}{A_1} \, .
\end{align}
Here $F^{(2)}$ and $A_1$ are positive constants.
If $F^{(2)}$ is small enough, the last term in \eqref{slFRV5} may give any contribution only at late time.

\section{Matter Contribution to Future Evolution of Universe\label{Sec5}}

We can include perfect matter fluids in our considerations. 
We further clarify how the matter behaves in the Jordan frame. 
The matter, of course, contributes to the future evolution of the Universe.

Varying the action~\eqref{JGRG21} with matter with respect to the metric, we obtain
\begin{align}
\label{Aeq}
\frac{1}{2}g_{\mu\nu} \left( F'(A)\left(R-A\right) + F(A) \right) - R_{\mu\nu} F'(A) - g_{\mu\nu} \Box F'(A) + \nabla_\mu \nabla_\nu F'(A) = - \frac{\kappa^2}{2}T_{\mathrm{matter}\, \mu\nu}\, ,
\end{align}
in addition to the equation given by the variation with respect to $A$,
\begin{align}
\label{Aeq2}
A=R\, .
\end{align}
By substituting \eqref{Aeq2} into \eqref{Aeq}, we obtain \eqref{JGRG13}. 
In the spatially flat FLRW Universe~\eqref{JGRG14}, the $(t,t)$ and $(i,j)$ components of \eqref{Aeq} have the following form,
\begin{align}
\label{Aeq3}
0=&\, - \frac{1}{2} \left( F'(A)\left(R-A\right) + F(A) \right) + 3 \left(\dot H + H^2 \right) F'(A) - 3H \frac{d}{dt} F'(A)  + \kappa^2 \rho_\mathrm{matter} \, , \nonumber \\
0 =&\, \frac{1}{2} \left( F'(A)\left(R-A\right) + F(A) \right) - \left( \dot H + 3 H^2 \right) F'(A) + \left( \frac{d^2}{dt^2} + 2 H \frac{d}{dt} \right) F'(A) + \kappa^2 p_\mathrm{matter} \, ,
\end{align}
with $A=R=12 H^2 + 6\dot H$.

We now consider the field-dependent Weyl rescaling in \eqref{JGRG22} by assuming that $\sigma$ only depends on the time coordinate.
Then the FLRW metric in \eqref{JGRG14} is transformed as,
\begin{align}
\label{JGRG14J}
ds^2 \to d{\tilde s}^2 = g_{\mu\nu}^\mathrm{Einstein\ frame} = - \e^{-\sigma(t)} dt^2 + \e^{-\sigma(t)} a(t)^2 \sum_{i=1,2,3} \left(dx^i\right)^2\, .
\end{align}
Then we define a new time-coordinate $\tilde t$ and the Hubble rate $\tilde H$ as follows,
\begin{align}
\label{dt}
d\tilde t = \e^{-\frac{\sigma(t)}{2}} dt \, , \quad
\tilde H = \frac{\e^{\frac{\sigma(t)}{2}}}{a(t)} \frac{d \left( \e^{-\frac{\sigma(t)}{2}} a(t) \right)}{d\tilde t}
= - \frac{1}{2} \frac{d\sigma}{d\tilde t} + \e^\frac{\sigma}{2} H \, ,
\end{align}
which gives,
\begin{align}
\label{dt2}
\frac{d}{dt} =&\, \e^{- \frac{\sigma}{2}} \frac{d}{d\tilde t} \, , \quad \frac{d^2}{dt^2} = \e^{-\sigma} \left( \frac{d^2}{d\tilde t^2} - \frac{1}{2} \frac{d\sigma}{d\tilde t} \frac{d}{d\tilde t} \right) \, , \nonumber \\
H =&\, \e^{-\frac{\sigma}{2}} \left( \tilde H + \frac{1}{2}\frac{d\sigma}{d\tilde t} \right)\, , \quad
\dot H = \e^{-\sigma} \left\{ - \frac{1}{2} \frac{d\sigma}{d\tilde t} \left( \tilde H + \frac{1}{2} \frac{d\sigma}{d\tilde t} \right)+ \frac{d\tilde H}{d\tilde t} + \frac{1}{2} \frac{d^2 \sigma}{d\tilde t^2} \right\} \, .
\end{align}
Then we find,
\begin{align}
\label{crvtrEin}
R = 6 \e^{-\sigma} \left( 2 {\tilde H}^2 + \frac{3}{2} \tilde H \frac{d\sigma}{d\tilde t} + \frac{1}{4}\left( \frac{d\sigma}{d\tilde t} \right)^2 + \frac{d\tilde H}{d\tilde t}
+ \frac{1}{2}\frac{d^2 \sigma}{d\tilde t^2} \right) \, .
\end{align}
Then the equations in \eqref{Aeq3} have the following forms,
\begin{align}
\label{Aeq3E}
0 =&\, \frac{1}{2} \e^{-2\sigma} V(\sigma) + 3 \e^{-2\sigma} \left( - {\tilde H}^2 + \frac{3}{4} \left( \frac{d\sigma}{d\tilde t} \right)^2 \right)
+ \kappa^2 \rho_\mathrm{matter} \, , \nonumber \\
0 =&\, - \frac{1}{2} \e^{-2\sigma} V(\sigma)
+ \e^{-2\sigma} \left( 3 {\tilde H}^2 + \frac{3}{4}\left( \frac{d\sigma}{d\tilde t} \right)^2
+ 2 \frac{d\tilde H}{d\tilde t} \right)
+ \kappa^2 p_\mathrm{matter} \, .
\end{align}
The above equations in \eqref{Aeq3E} are nothing but the Friedmann equations for the action~\eqref{JGRG23} in the Einstein frame,
\begin{align}
\label{Aeq3EFrdmnn}
3 {\tilde H}^2=&\, \frac{3}{4} \left( \frac{d\sigma}{d\tilde t} \right)^2 + \frac{1}{2} V(\sigma) + \kappa^2 \e^{2\sigma} \rho_\mathrm{matter} \, , \nonumber \\
- \left( 3 {\tilde H}^2 + 2 \frac{d\tilde H}{d\tilde t} \right) = &\, \frac{3}{4}\left( \frac{d\sigma}{d\tilde t} \right)^2 - \frac{1}{2} V(\sigma) + \kappa^2  \e^{2\sigma} p_\mathrm{matter} \, .
\end{align}
Then we find the effective transformation law between the Jordan frame and the Einstein frame,
\begin{align}
\label{JErhp}
\rho_\mathrm{matter} \to \tilde\rho_\mathrm{matter} = \e^{2\sigma} \rho_\mathrm{matter} \, , \quad
p_\mathrm{matter} \to \tilde p_\mathrm{matter} = \e^{2\sigma} p_\mathrm{matter} \, .
\end{align}
On the other hand, the equation $A=R=12 H^2 + 6\dot H$ has the following form,
\begin{align}
\label{sR}
A = 6 \e^{-\sigma} \left( 2 {\tilde H}^2 + \frac{3}{2} \tilde H \frac{d\sigma}{d\tilde t} + \frac{1}{4}\left( \frac{d\sigma}{d\tilde t} \right)^2 + \frac{d\tilde H}{d\tilde t}
+ \frac{1}{2}\frac{d^2 \sigma}{d\tilde t^2} \right) \, .
\end{align}
The conservation of matter~\eqref{cnsv} has the following form,
\begin{align}
\label{cnsvJ}
0 = \e^{-\frac{\sigma}{2}} \left\{ \frac{d\rho_\mathrm{matter}}{d\tilde t} + 3 \left( \tilde H + \frac{1}{2}\frac{d\sigma}{d\tilde t} \right) \left( \rho_\mathrm{matter} + p_\mathrm{matter} \right) \right\}\, .
\end{align}
When the EoS parameter of the matter is a constant $w$, $p_\mathrm{matter} = w \rho_\mathrm{matter}$, Eq.~\eqref{cnsvJ} can be integrated to,
\begin{align}
\label{rhoE}
\rho_\mathrm{matter} = \rho_\mathrm{matter}^{(0)} {\tilde a}^{-3(1+w)} \e^{- \frac{3(1+w)}{2} \sigma} \, .
\end{align}
Here $\tilde a$ is the scale factor in the Einstein frame, $\tilde a= \e^{- \frac{\sigma}{2}} a$. 
If we use the relation, we obtain a standard expression of  $\rho_\mathrm{matter}$ in the Jordan frame, 
\begin{align}
\label{rhoJ}
\rho_\mathrm{matter} = \rho_\mathrm{matter}^{(0)} a^{-3(1+w)} \, .
\end{align}
By combining the two equations in \eqref{Aeq3EFrdmnn}, we obtain,
\begin{align}
\label{cmbn}
\frac{d\tilde H}{d\tilde t} = - \frac{3}{4}\left( \frac{d\sigma}{d\tilde t} \right)^2 - \frac{\kappa^2}{2} \e^{2\sigma} \left( \rho_\mathrm{matter} + p_\mathrm{matter}\right)  \, .
\end{align}
By using the first equation in \eqref{Aeq3EFrdmnn} and Eq.~\eqref{cmbn}, we may delete ${\tilde H}^2$ and $\frac{d\tilde H}{d\tilde t}$ in \eqref{sR} and obtain,
\begin{align}
\label{sRB}
0  =&\, \frac{1}{6}\left\{ 3 \left( \frac{d^2 \sigma}{d\tilde t^2} + 3 \tilde H \frac{d\sigma}{d\tilde t} \right) + \frac{dV(\sigma)}{d\sigma}
 - \kappa^2 \e^{2\sigma} \left( - \rho_\mathrm{matter} + 3 p_\mathrm{matter}\right) \right\}  \, .
\end{align}
Here we have used \eqref{FRV1B}. The equation is nothing but the scalar field equation in the Einstein frame action~\eqref{JGRG23}.
The last term, given by the trace of the matter energy-momentum tensor, comes from the coupling of $\sigma$ and matter via the field-dependent Weyl rescaling~\eqref{JGRG22}, as expected. 
Eq.~\ref{sRB} gives the evolution of the scalar field derived by the combination of the potential $V(\sigma)$ and the coupling with matter.

Eq.~\eqref{sRB} also tells us that the effective potential including the contribution from matter is given by
\begin{align}
\label{eff}
V_\mathrm{eff} (\sigma) = V(\sigma) - \frac{\kappa^2}{2} \e^{2\sigma} \left( - \rho_\mathrm{matter} + 3 p_\mathrm{matter}\right) \, .
\end{align}
In the case of $F(R)=f_0 R^m$, we find
\begin{align}
\label{sigmaA}
\e^{-\sigma} = mf_0 A^{m-1}\quad \mbox{or} \quad A = \left( m f_0 \right)^{-\frac{1}{m-1}} \e^{- \frac{\sigma}{m-1}}\, ,
\end{align}
and therefore the potential $V(\sigma)$ in \eqref{VFpwr} has the following form,
\begin{align}
\label{VFpwr2}
V (\sigma) = \frac{1}{m f_0} \left( 1 -  \frac{1}{m}\right) \left( m f_0 \right)^\frac{m-2}{m-1} \e^{\frac{m-2}{m-1}\sigma} \, .
\end{align}
Then the minimum of the effective potential is given by solving the following equation,
\begin{align}
\label{VFpwrdsh}
V_\mathrm{eff}' (\sigma)  = \frac{m-2}{m} \left( m f_0 \right)^{ - \frac{1}{m-1}} \e^{\frac{m-2}{m-1}\sigma}
- \kappa^2 \e^{2\sigma} \left( - \rho_\mathrm{matter} + 3 p_\mathrm{matter}\right) \, ,
\end{align}
whose solution is,
\begin{align}
\label{VFpwrdshsl}
\sigma = - \frac{m-1}{m} \ln \left( \frac{m \left( m f_0 \right)^\frac{1}{m-1} \kappa^2 \left( - \rho_\mathrm{matter} + 3 p_\mathrm{matter}\right)}{m-2} \right)\, .
\end{align}
In order that $\sigma$ in \eqref{VFpwrdshsl} is a real number solution, we require,
\begin{align}
\label{sigmasol}
\frac{m \left( - \rho_\mathrm{matter} + 3 p_\mathrm{matter}\right)}{m-2} > 0 \, .
\end{align}
Here we have used the condition~\eqref{Fpwrd2}. 
If the evolution of the spacetime is slow, that is, the evolution of $\rho_\mathrm{matter}$ and $p_\mathrm{matter}$ is small, $\sigma$ could sit at the minimum. 
Because the potential $V(\sigma)$ in \eqref{VFpwr2} is given by the model~\eqref{Fpwr}, by using \eqref{sigmaA},
\begin{align}
\label{A0}
A = \left( m f_0 \right)^{-\frac{1}{m-1}} \left( \frac{m \left( m f_0 \right)^\frac{1}{m-1} \kappa^2 \left( - \rho_\mathrm{matter} + 3 p_\mathrm{matter}\right)}{m-2} \right)^\frac{1}{m} \, .
\end{align}
Because $A$ is the scalar curvature in the Jordan frame $A=R=12 H^2 + 6\dot H$, Eq.~\eqref{A} gives a differential equation,
\begin{align}
\label{A}
12 H^2 + 6\dot H = \left( m f_0 \right)^{-\frac{1}{m-1}} \left( \frac{m \left( m f_0 \right)^\frac{1}{m-1} \kappa^2 \left( - \rho_\mathrm{matter} + 3 p_\mathrm{matter}\right)}{m-2} \right)^\frac{1}{m} \, .
\end{align}
Especially when the EoS parameter $w$of the matter is a constant, by using \eqref{rhoJ}, Eq.~\eqref{A} has the following form,
\begin{align}
\label{A2}
12 H^2 + 6\dot H = \left( m f_0 \right)^{-\frac{1}{m-1}} \left( \frac{m \left( m f_0 \right)^\frac{1}{m-1} \kappa^2 \left( 3w - 1 \right) \rho_\mathrm{matter}^{(0)} a^{- 3(1+w)} }{m-2} \right)^\frac{1}{m} \, .
\end{align}
If we assume,
\begin{align}
\label{sola}
a = \left( \frac{t_0}{t} \right)^{-\frac{2m}{3(1+w)}} \, ,
\end{align}
which gives,
\begin{align}
\label{solH}
H = \frac{\frac{2m}{3(1+w)}}{t}\, ,
\end{align}
Eq.~\eqref{A2} reduces to an algebraic equation,
\begin{align}
\label{aleq}
\frac{16m^2}{3(1+w)^2} + \frac{4m}{1+w} = \left( m f_0 \right)^{-\frac{1}{m-1}} \left( \frac{m \left( m f_0 \right)^\frac{1}{m-1} \kappa^2 \left( 3w - 1 \right) \rho_\mathrm{matter}^{(0)} }{m-2} \right)^\frac{1}{m} {t_0}^2 \, ,
\end{align}
which determines $t_0$ as follows,
\begin{align}
\label{t0}
{t_0}^2 = \frac{\frac{16m^2}{3(1+w)^2} + \frac{4m}{1+w}}{ \left( m f_0 \right)^{-\frac{1}{m-1}} \left( \frac{m \left( m f_0 \right)^\frac{1}{m-1} \kappa^2 \left( 3w - 1 \right) \rho_\mathrm{matter}^{(0)} }{m-2} \right)^\frac{1}{m}} \, .
\end{align}
Eq.~\eqref{solH} is consistent with \eqref{M8}. 
For the model~\eqref{model1} with $m_e>1$ and $m_l<1$, by using \eqref{solH}, the behavior of $H$ in the late-time is given by, 
\begin{align}
\label{solHB}
H = \frac{\frac{2m_l}{3(1+w)}}{t}\, .
\end{align}
We may assume that dust with $w=0$ dominates in the late Universe.
Then Eq.~\eqref{t0} becomes,
\begin{align}
\label{t01}
{t_0}^2 = \frac{\frac{16m_l\left(m_l + \frac{3}{4} \right)}{3} }{ \left( m_l f_0 \right)^{-\frac{1}{m_l-1}} \left( - \frac{m_l \left( m_l f_0 \right)^\frac{1}{m_l-1} \kappa^2 \rho_\mathrm{matter}^{(0)} }{m_l -2} \right)^\frac{1}{m_l}} \, .
\end{align}
Then, so that ${t_0}^2$ is real and positive, we require $1<m_l<2$.

\section{Future Evolution in Realistic Models\label{Sec6}}

In this section, we discuss the future evolution of the Universe in some realistic models proposed in \cite{Odintsov:2026fyc}. 
We show that the anti-gravity or ghost graviton often appears in the future for these models.

In \cite{Odintsov:2026fyc}, as realistic models of $F(R)$ gravity, the model including a logarithmic function and the model including an exponential function have been considered,
\begin{align}
\label{lnmodel}
F(R)=&\, R+\frac{R^2}{M^2} - 2\Lambda\left\{ 1-\frac{1}{1+\gamma\ln \left(\frac{\alpha R}{2\Lambda}\right)}\right\} \, , \\
\label{expmodel}
F(R)=&\, \frac{R^2}{M^2} +R\exp\left\{ \varepsilon\left(\frac{2\Lambda}{R}\right)^\beta\right\} +\nu\Lambda \, .
\end{align}
Here $M$, $\Lambda$, $\gamma$, $\alpha$, $\varepsilon$, $\beta$, and $\nu$ are constants. 
The parameters $M$, $\Lambda$, $\gamma$, $\alpha$, $\varepsilon$ are positive and $\beta\geq 1$ and $\nu$ is given by,
\begin{align}
\label{nu}
\nu=\left\{ \begin{array}{ll}
-2\, ,&\beta>1,\\
-2-2\varepsilon\, ,&\beta=1
\end{array} \right. .
\end{align}
In order to generate inflation, the term $\frac{R^2}{M^2}$ corresponding to the Starobinsky model~\cite{Starobinsky:1980te} is included.

Because for the logarithmic model,
\begin{align}
\label{Fdrvln}
F'(R) =&\, 1 + \frac{2R}{M^2} - \frac{\frac{2\Lambda\gamma}{R}}{\left( 1+\gamma\ln \left(\frac{\alpha R}{2\Lambda}\right)\right)^2}\, , \nonumber \\
F''(R) =&\, \frac{2}{M^2} + \frac{\frac{2\Lambda\gamma}{R^2}}{\left( 1+\gamma\ln \left(\frac{\alpha R}{2\Lambda}\right)\right)^2}
+ \frac{\frac{4\Lambda\gamma^2}{R^2}}{\left( 1+\gamma\ln \left(\frac{\alpha R}{2\Lambda}\right)\right)^3}\, ,
\end{align}
and for the exponential model,
\begin{align}
\label{Fdrvexp}
F'(R) =&\, \frac{2R}{M^2} + \left\{ 1 - \varepsilon\beta \left(\frac{2\Lambda}{R}\right)^\beta \right\} \exp\left\{\varepsilon\left(\frac{2\Lambda}{R}\right)^\beta\right\}
\, , \nonumber \\
F''(R) =&\, \frac{2}{M^2} + \left\{ \frac{\beta\left(\beta - 1\right) \varepsilon}{R} \left(\frac{2\Lambda}{R}\right)^\beta
+ \frac{\beta^2\varepsilon^2}{R} \left(\frac{2\Lambda}{R}\right)^{2\beta} \right\} \exp\left\{\varepsilon\left(\frac{2\Lambda}{R}\right)^\beta\right\}
\, , \nonumber \\
\end{align}
we obtain the potential $V$ in \eqref{JGRG23} for the logarithmic model has the following form,
\begin{align}
\label{JGRG23ln}
V(A) = \frac{\frac{A^2}{M^2} - \frac{2\Lambda\gamma}{\left( 1+\gamma\ln \left(\frac{\alpha A}{2\Lambda}\right)\right)^2}
+ 2\Lambda\left[1-\frac{1}{1+\gamma\ln \left(\frac{\alpha A}{2\Lambda}\right)}\right]}
{ \left\{ 1 + \frac{2A}{M^2} - \frac{\frac{2\Lambda\gamma}{A}}{\left( 1+\gamma\ln \left(\frac{\alpha A}{2\Lambda}\right)\right)^2} \right\}^2} \, ,
\end{align}
and the potential for the exponential model,
\begin{align}
\label{JGRG23exp}
V(A) = \frac{\frac{A^2}{M^2} - \varepsilon\beta A^2 \left(\frac{2\Lambda}{A}\right)^\beta \exp\left\{\varepsilon\left(\frac{2\Lambda}{A}\right)^\beta \right\} - \nu\Lambda}
{\left[ \frac{2A}{M^2} + \left\{ 1 - \varepsilon\beta \left(\frac{2\Lambda}{A}\right)^\beta \right\} \exp\left\{\varepsilon\left(\frac{2\Lambda}{A}\right)^\beta \right\}\right]^2} \, .
\end{align}
On the other hand, the extrema for the logarithmic model are given by the solutions of the equation corresponding to \eqref{FRV2},
\begin{align}
\label{FRV2ln}
0 = A + \frac{2\Lambda\gamma}{\left( 1+\gamma\ln \left(\frac{\alpha A}{2\Lambda}\right)\right)^2}
 - 4\Lambda\left\{ 1-\frac{1}{1+\gamma\ln \left(\frac{\alpha A}{2\Lambda}\right)}\right\} \, ,
\end{align}
and the extrema for the exponential model are given by,
\begin{align}
\label{FRV2exp}
0= A \left\{ 1 + \varepsilon\beta \left(\frac{2\Lambda}{A}\right)^\beta \right\} \exp\left\{\varepsilon\left(\frac{2\Lambda}{A}\right)^\beta\right\}
+2\nu\Lambda \, .
\end{align}
In the following, we consider the possibility that a stable de~Sitter spacetime could be realized.

\subsection{Logarithmic model}

First, we consider the logarithmic model \eqref{lnmodel} in more detail.
As clear from \eqref{lnmodel}, $F(R)$ has a singularity when $1+\gamma\ln \left(\frac{\alpha A}{2\Lambda}\right)$ vanishes, that is
\begin{align}
\label{Rc}
R=R_c \equiv \frac{2\Lambda}{\alpha} \e^{- \frac{1}{\gamma}} \, .
\end{align}
When $R\sim R_c$, Eq.~\eqref{Fdrvln} tells us that $F'(R)$ becomes negative and therefore the anti-gravity and a ghost graviton appear, and the model becomes inconsistent. 
In order to avoid the problem, the Universe goes to a stable de~Sitter vacuum before reaching the region where $F'(R)$ is negative. 
The de~Sitter Universe is given by the solution $A=R$ in \eqref{FRV2ln}. 
We rewrite \eqref{FRV2ln} as follows,
\begin{align}
\label{FRV2ln2}
 -A  + 2 \Lambda \left( 2 + \frac{1}{\gamma} \right) =&\, 2\Lambda\gamma \left(\frac{1}{1+\gamma\ln \left(\frac{\alpha A}{2\Lambda}\right)} + \frac{1}{\gamma} \right)^2 \, .
\end{align}
We should note that when $A=R\to R_c$ but $A=R>R_c$, $\frac{1}{1+\gamma\ln \left(\frac{\alpha A}{2\Lambda}\right)} \to + \infty$ but when $A=R\to R_c$ but $A=R<R_c$, $\frac{1}{1+\gamma\ln \left(\frac{\alpha A}{2\Lambda}\right)} \to - \infty$. 
When $A=R\to 0$ or $A=R\to \infty$, we find $\frac{1}{1+\gamma\ln \left(\frac{\alpha A}{2\Lambda}\right)} \to 0$. 
Therefore, when $+\infty>R>R_c$, $\frac{1}{1+\gamma\ln \left(\frac{\alpha A}{2\Lambda}\right)}>0$ and when $R_c>R>0$, $\frac{1}{1+\gamma\ln \left(\frac{\alpha A}{2\Lambda}\right)}<0$.

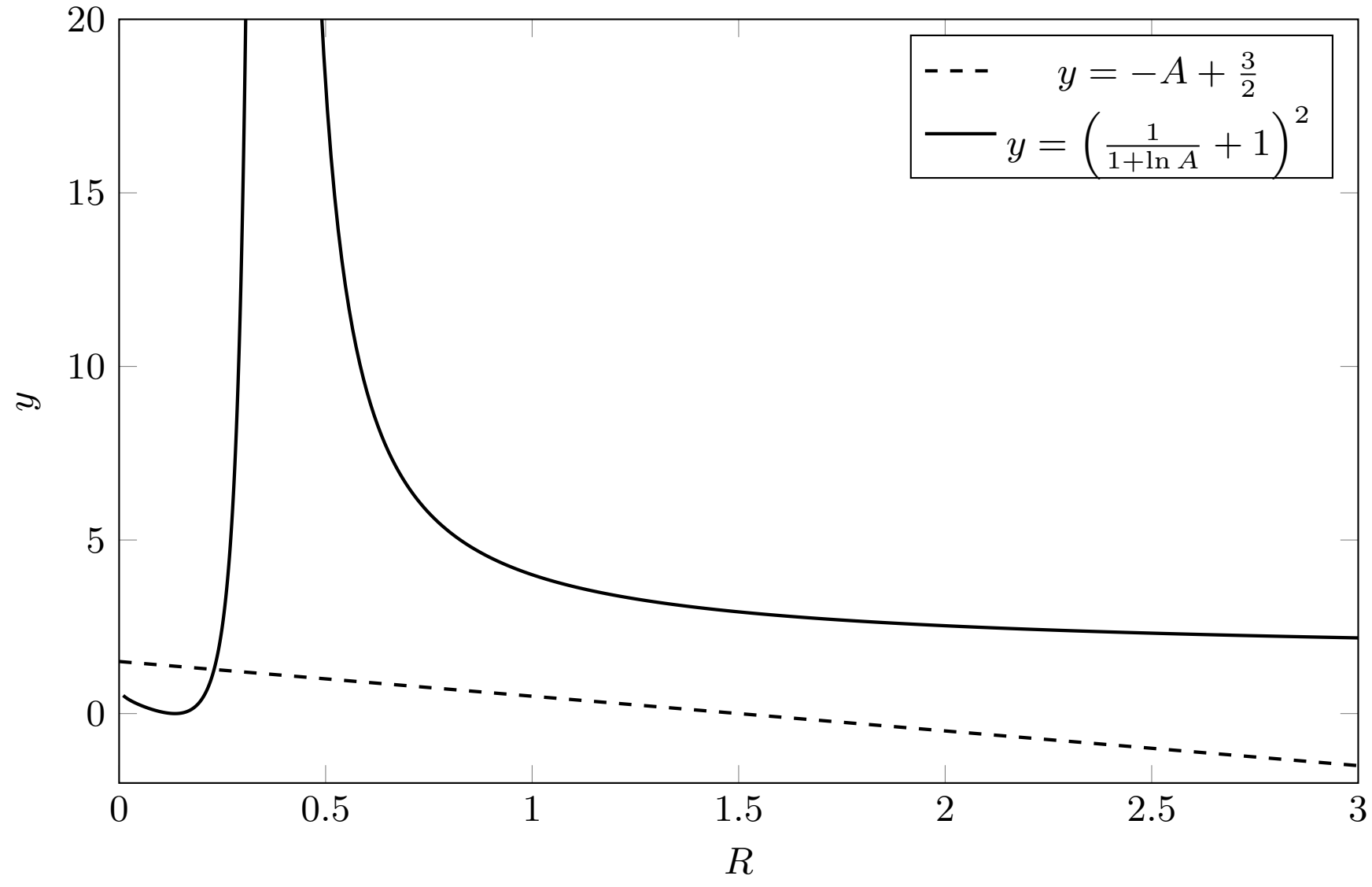
\begin{figure}[h]
\begin{center}

\begin{tikzpicture}
\begin{axis}[
width=12cm,
height=8cm,
xmin=0,
xmax=3,
ymin=-2,
ymax=20,
restrict y to domain=-2:20
axis lines=middle,
xlabel={$R$},
ylabel={$y$},
samples=400,
legend style={
at={(0.98,0.98)},
anchor=north east,
},
]

\addplot[
black,
dashed,
thick,
domain=0:3 ]
{-x + 1.5};
\addlegendentry{$y=-A+\frac32$}

\addplot[
black,
thick,
domain=0.01:0.34,
samples=300
]
{(1/(1+ln(x))+1)^2};

\addplot[
black,
thick,
domain=0.40:3,
samples=300
]
{(1/(1+ln(x))+1)^2};
\addlegendentry{$y=\left(\frac{1}{1+\ln A}+1\right)^2$}

\end{axis}
\end{tikzpicture}

\end{center}

\caption{
The behaviors of the l.h.s. of \eqref{FRV2ln2} (dashed line) and the r.h.s. (solid line).
Here we have chosen $2 \Lambda \left( 2 + \frac{1}{\gamma} \right) =\frac{3}{2}$, $2\Lambda\gamma=1$, $\frac{\alpha}{2\Lambda}=1$, and $\gamma=1$.
The point of intersection of a line and a curve s corresponds to the de~Sitter spacetime, which appears in the region of $0<A<R_c=\frac{1}{\e}$.
}
\label{dSpoint}
\end{figure}

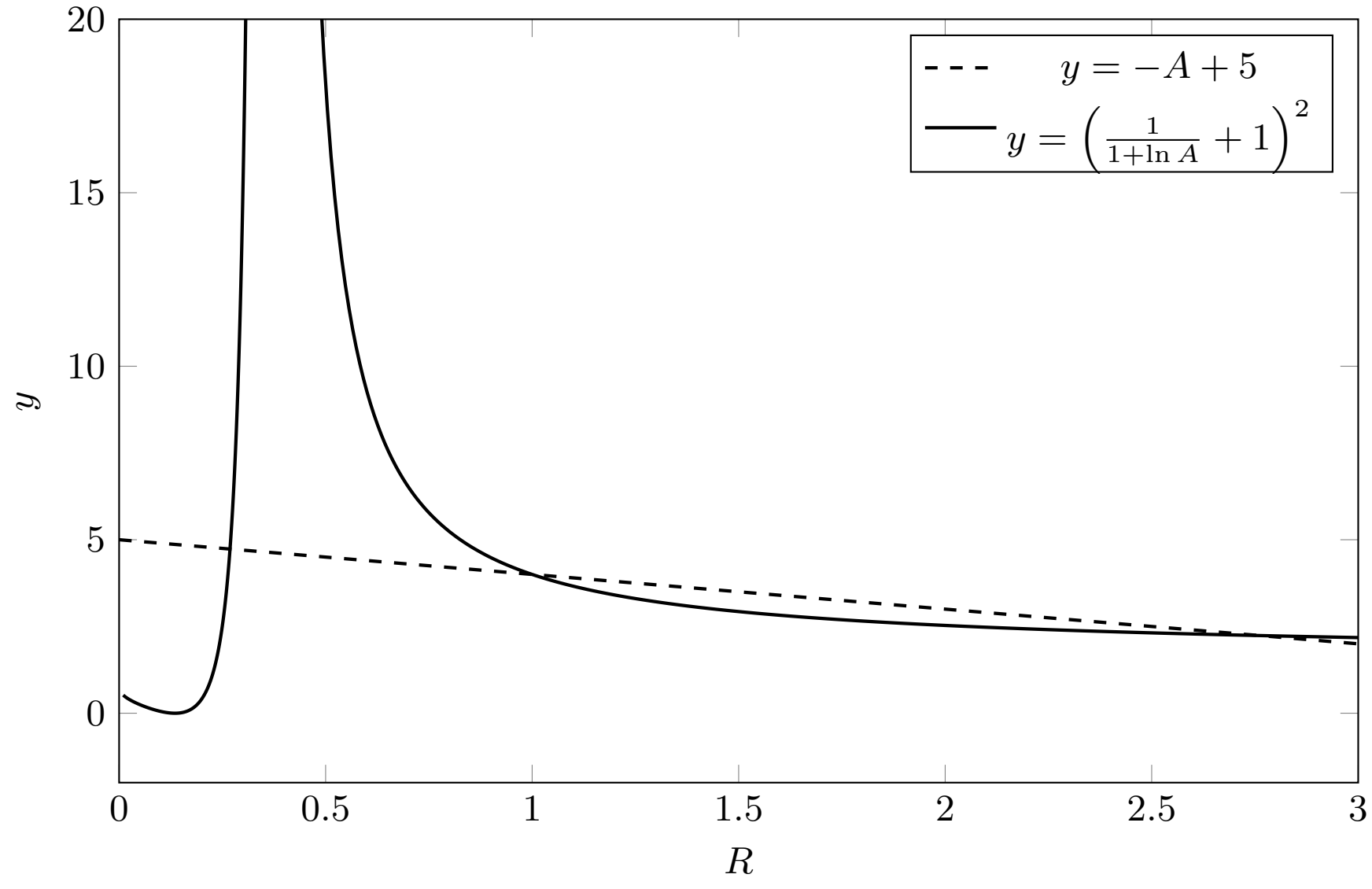
\begin{figure}[h]
\begin{center}

\begin{tikzpicture}
\begin{axis}[
width=12cm,
height=8cm,
xmin=0,
xmax=3,
ymin=-2,
ymax=20,
restrict y to domain=-2:20
axis lines=middle,
xlabel={$R$},
ylabel={$y$},
samples=400,
legend style={
at={(0.98,0.98)},
anchor=north east,
},
]

\addplot[
black,
dashed,
thick,
domain=0:3 ]
{-x + 5};
\addlegendentry{$y=-A+5$}

\addplot[
black,
thick,
domain=0.01:0.34,
samples=300
]
{(1/(1+ln(x))+1)^2};

\addplot[
black,
thick,
domain=0.40:3,
samples=300
]
{(1/(1+ln(x))+1)^2};
\addlegendentry{$y=\left(\frac{1}{1+\ln A}+1\right)^2$}

\end{axis}
\end{tikzpicture}

\end{center}

\caption{ The behaviors of the l.h.s. of \eqref{FRV2ln2} (dashed line) and the r.h.s. (solid line). 
Here we have chosen $2 \Lambda \left( 2 + \frac{1}{\gamma} \right) =5$, $2\Lambda\gamma=1$, $\frac{\alpha}{2\Lambda}=1$, and $\gamma=1$. 
The points of intersection of a line and a curve correspond to the de~Sitter spacetime, one of which appears in the region of $0<A<R_c=\frac{1}{\e}$ and the other two appear in the region of $R_c<A<+\infty$. 
The de~Sitter space corresponding to the larger solution of $A$ in the region of $R_c<A<+\infty$ is stable. }
\label{dSpoint2}
\end{figure}

When $A=R\to 0$, the l.h.s. in Eq.~\eqref{FRV2ln2} becomes $2 \Lambda \left( 2 + \frac{1}{\gamma} \right)$ and the r.h.s. becomes $\frac{2\Lambda}{\gamma}$, and therefore the l.h.s. is
larger than the r.h.s. 
When $A=R\to R_c$, the l.h.s. in Eq.~\eqref{FRV2ln2} becomes a finite value $- R_c + 2 \Lambda \left( 2 + \frac{1}{\gamma} \right)$ but the r.h.s. becomes $+\infty$. 
When $A=R\to +\infty$, the l.h.s. in Eq.~\eqref{FRV2ln2} becomes $- \infty$ and the r.h.s. becomes $\frac{2\Lambda}{\gamma}$. 
Therefore, there is at least one solution of Eq.~\eqref{FRV2ln2} between $0<R<R_c$. 
The r.h.s. in Eq.~\eqref{FRV2ln2} vanishes when $\frac{1}{1+\gamma\ln \left(\frac{\alpha A}{2\Lambda}\right)} = - \frac{1}{\gamma}<0$.
Because $-\frac{1}{\gamma}$ is negative, the corresponding value of $R$ exists in $0<R<R_c$. 
The behaviors of the l.h.s. of \eqref{FRV2ln2} and the r.h.s. are shown in FIG.~\ref{dSpoint}. 
When $\gamma$ is not so small, as in FIG.~\ref{dSpoint}, there is only one solution, but as in FIG.~\ref{dSpoint2}, when $\gamma$ is small enough, there appear three solutions.

When $R=A$ is large, in the potential~\eqref{JGRG23ln}, the terms coming from the Starobinsky model becomes dominant, $V(A)\sim \frac{\frac{A^2}{M^2}}{\left(1 + \frac{2A}{M^2} \right)^2}$, which is a monotonically increasing function of $A$. 
When $\lambda$ is not so small as in FIG.~\ref{dSpoint}, because there is no de~Sitter solution or no extremum of the full potential~\eqref{JGRG23ln} during $+\infty > R > R_c$, the full potential is also a monotonically increasing function of $A$.
Because $A$ reaches the region where $F'(A)<0$ before $A$ reaches $A=R_c$, the model shows anti-gravity and the ghost graviton appears in a finite future. 
On the other hand, when $\lambda$ is small enough as in FIG.~\ref{dSpoint2}, there are three solutions corresponding to the points of intersection of a line and a curve, which correspond to the de~Sitter spacetime, one of which appears in the region of $0<A<R_c=\frac{1}{\e}$ and the other two appear in the region of $R_c<A<+\infty$. 
The de~Sitter space corresponding to the larger solution of $A$ in the region of $R_c<A<+\infty$ is stable because in the region larger than the larger solution, the potential is a monotonically increasing function of $A$. 
If a proper initial condition is chosen, the spacetime might not go to the anti-gravity region where $F'(A)<0$, but the spacetime might stop at the stable de~Sitter spacetime.

In \cite{Odintsov:2026fyc}, the best-fit parameters from the observational data are,
\begin{align}
\label{lnftprmtr}
\alpha = 0.80^{+1.60}_{-0.794} \, , \quad
\gamma = 1.483^{+1.565}_{-1.293} \, .
\end{align}
Let $\alpha = 0.80$ and $\gamma = 1.483$.
Then the function $F(R)$ in \eqref{lnmodel} has a pole at
\begin{align}
\label{lnFRpl}
\frac{R}{2\Lambda} = \frac{\e^{-\frac{1}{\gamma}}}{\alpha} \sim 0.6369\, .
\end{align}
On the other hand, when we neglect the term $\frac{2R}{M^2}$ in the first equation of \eqref{Fdrvln} because the term is only relevant when the curvature is large, we find that $F'(R)=0$ has
the following solution, 
\begin{align}
\label{Fdzeroln}
\frac{R}{2\Lambda} \sim 1.3064 \, .
\end{align}
Equation $F'(R)=0$ has two solutions, and the solution in \eqref{Fdzeroln} is the larger one because we are considering the decreasing curvature from the large curvature region. 
Then, for the Universe to go to the asymptotically de~Sitter spacetime before the Universe goes to the anti-gravity or ghost graviton region, Eq.~\eqref{FRV2ln} must have a solution when $\frac{R}{2\Lambda} > 1.3064$. 
We find, however, that there is no solution in \eqref{FRV2ln} when $\frac{R}{2\Lambda} > 1.3064$, as in FIG.~\ref{dSpoint}. 
Therefore, the realistic model always gives the unphysical anti-gravity or ghost graviton region in the future. 
We should note, however, that we can add some terms which dominate only in small curvature but before the curvature reaches the anti-gravity or ghost graviton region, that is, when $\frac{R}{2\Lambda} > 1.3064$ as found in \eqref{Fdzeroln}. 
For example, we may modify the model so that the Universe goes to the phantom phase in the future, that is, the curvature has a minimum and does not reach $\frac{R}{2\Lambda} \sim 1.3064$ in \eqref{Fdzeroln}. 
Another but general prescription to solve the problem is given in the next section.

\subsection{Exponential model}

Now we consider the exponential model \eqref{expmodel} in more detail. 
As is clear from \eqref{Fdrvexp}, when $R$ is small, $F'(R)$ behaves as,
\begin{align}
\label{Fdrvexp2}
F'(R)\sim &\, - \varepsilon\beta \left(\frac{2\Lambda}{R}\right)^\beta \exp\left\{\varepsilon\left(\frac{2\Lambda}{R}\right)^\beta\right\} <0 \, , \nonumber \\
\end{align}
and therefore $F'(R)$ becomes negative, leading to anti-gravity and the ghost graviton. 
A possible scenario could be that before the scalar curvature $R$ arrives at the region where $F'(R)$ is negative, there is a stable solution describing the de~Sitter spacetime.

In the case $\beta=1$, the situation becomes rather simple. 
The condition \eqref{FRV2exp} for the de~Sitter spacetime has the following form,
\begin{align}
\label{FRV2expbeta1}
0 =&\, \frac{A}{2\Lambda} \left\{ 1 + \varepsilon \left(\frac{2\Lambda}{A}\right) \right\} \exp\left\{\varepsilon\left(\frac{2\Lambda}{A}\right) \right\}
 - 2\left( 1 + \varepsilon\right) \, .
\end{align}
Here we used \eqref{nu}. If $\varepsilon$ is small enough, we find,
\begin{align}
\label{exp2}
A = R \sim 4 \Lambda \, .
\end{align}
Then, by assuming $\Lambda \ll M$, the equations in \eqref{Fdrvexp} tell,
\begin{align}
\label{Fdrvexpbeta1}
F'(R) \sim 1 \, , \quad
F''(R) = \frac{2}{M^2} + \frac{\varepsilon^2}{R} \left(\frac{2\Lambda}{R}\right)^2 \exp\left\{\varepsilon\left(\frac{2\Lambda}{R}\right)\right\} > 0 \quad
\Rightarrow \quad F''(R) \sim \frac{2}{M^2} + \frac{\varepsilon^2}{16\Lambda} \, ,
\end{align}
and we obtain
\begin{align}
\label{FRV4beta1}
 - A + \frac{F'(A)}{F''(A)} \sim - 4\Lambda + \frac{\Lambda}{\frac{2\Lambda}{M^2} + \frac{\varepsilon^2}{16}} \sim \frac{\Lambda}{\frac{2\Lambda}{M^2} + \frac{\varepsilon^2}{16}} > 0\, .
\end{align}
Therefore, Eq.~\eqref{FRV4} is satisfied for the solution~\eqref{exp2} and the corresponding de~Sitter spacetime is stable. 
If a proper initial condition is chosen, the spacetime might not go to the anti-gravity region where $F'(A)<0$, but the spacetime might stop at the stable de~Sitter spacetime.

Because for the exponential model~\eqref{expmodel}, the best-fit parameters from the observational data are given by \cite{Odintsov:2026fyc},
\begin{align}
\label{bstexp}
\beta = 1.00^{+0.36}_{-0.00}\, , \quad
\varepsilon = 1.02^{+2.52}_{-0.75} \, ,
\end{align}
we consider the case $\beta=1$ and $\varepsilon= 1.02$, which gives $\nu=-4.04$ by using \eqref{nu}. 
The condition of the de~Sitter spacetime \eqref{FRV2exp} has the following form, 
\begin{align}
\label{dSexp}
\frac{A}{2\Lambda} \exp \left\{ \frac{1.02}{\left( \frac{A}{2\Lambda} \right)} \right\} \left\{ 1 + \frac{1.02}{\left( \frac{A}{2\Lambda} \right)} \right\} = 4.04\, .
\end{align}
The minimum value of the l.h.s. of \eqref{dSexp} is given when $\frac{A}{2\Lambda} \sim 1.6504$ as $\frac{A}{2\Lambda} \exp \left\{ \frac{1.02}{\left( \frac{A}{2\Lambda} \right)} \right\} \left\{ 1 + \frac{1.02}{\left( \frac{A}{2\Lambda} \right)} \right\} \sim 4.9563>4.04$. 
Therefore, Eq.~\eqref{FRV2exp} or \eqref{dSexp} has no solution. 
This tells, again, that the realistic model always gives the anti-gravity or ghost graviton region in the future, which is unphysical.

As in the case of the logarithmic model~\eqref{lnmodel}, we may add some terms which dominate only in small curvature but before the curvature reaches the anti-gravity or ghost graviton region. 
Such a term may make the curvature turn to increase, but a prescription which can be applied to general models is given in the next section.

\subsection{Phenomena before going to the anti-gravity region}

As discussed around Eq.~\eqref{FR1}, and as is clear from \eqref{JGRG21}, anti-gravity implies that the graviton is a ghost in quantum theory. 
Because the ghost conflicts with quantum theory, the theory including ghosts cannot be physically accepted.
For both the logarithmic and exponential models discussed in this section, although they inevitably evolve into the anti-gravity or de Sitter ghost-graviton region, as mentioned, the next section gives the prescription to avoid the problem for the general model. 
It might be, however, interesting to consider what could happen when the Universe approaches the anti-gravity region.

We should also note that we need to reconsider the Friedmann equation~\eqref{JGRG15}. 
When $F'(R) \sim 0$ near the transition point to the anti-gravity region, the Friedmann equation~\eqref{JGRG15} is not singular,
\begin{align}
\label{JGRG15ag}
0 \sim -\frac{F(R)}{2} - 18 \left( 4H^2 \dot H + H \ddot H\right) F''(R) + \kappa^2 \rho_\mathrm{matter}\, .
\end{align}
The contribution from the matter could be neglected in the late Universe, $\rho_\mathrm{matter} \sim 0$. 
Therefore, $H$ could not have any singularity, and therefore the Universe goes into the anti-gravity region in a finite time, at least classically.

Because Newton's gravitational constant $G$ is given by $\kappa^2 = 8\pi G$, Eq.~\eqref{JGRG21} tells that the effective gravitational constant $G_\mathrm{eff}$ is given by,
\begin{align}
\label{effG}
G_\mathrm{eff} = \frac{G}{F'(A)}\, .
\end{align}
If we can use Newton's gravity, when $F'(A)$ is positive but very small, $F'(A) \to 0+$ because the Universe approaches the anti-gravity region, the gravitational force becomes very strong and any objects could collapse. 
We should note, however, that when the effective gravitational constant $G_\mathrm{eff}$ is very large, non-linear effects and/or quantum effects will become very strong and therefore Newton's gravity will not be valid. 
Such non-linear effects and/or quantum effects may also work to prevent the Universe from going to the anti-gravity region.

Another interesting phenomenon might happen for the black holes.
Because now the horizon radius $r_\mathrm{horizon}$ of a black hole with mass $M$ is given by,
\begin{align}
\label{rhrzn}
r_\mathrm{horizon} = 2G_\mathrm{eff} M = \frac{2GM}{F'(A)}\, ,
\end{align}
the radius becomes very large near the transition to the anti-gravity region, and therefore the Universe might be swallowed up by black holes.

\section{Avoiding the anti-gravity\label{Sec7}}

In this section, we propose a mechanism using the potential barrier of the scalar field so that we can avoid the problem of the anti-gravity or ghost graviton in the realistic models in the last section.

In general, in order to avoid the anti-gravity or ghost graviton region, we need to fine-tune the parameters and the initial conditions; even if there is a stable de~Sitter point, the scalar field or $A=R$ goes through the de~Sitter point to reach the unphysical region. 
If there is a potential barrier, that is, the potential goes to positive infinity before reaching the unphysical region.

An example is given by, 
\begin{align}
\label{md}
F(R) \to \tilde F(R) \equiv F(R) \e^{ - \frac{{R_1}^2}{\left( R - R_0 \right)^2}}\, .
\end{align}
Here we choose $R_0$ so that $R_0$ is smaller than the value of the curvature in the unphysical anti-gravity or ghost gravity region. 
We also choose the parameter $R_1$ much smaller than the value of the scalar curvature in the present Universe. 
Therefore, $\e^{ - \frac{{R_1}^2}{\left( R_0 - R \right)^2}}\sim 1$ except in the region $R\sim R_0$, and therefore the properties of the original $F(R)$ do not change from the early Universe to the present Universe. 
We should also note that as long as $R>R_0$, the function $\e^{ - \frac{{R_1}^2}{\left( R_0 - R \right)^2}}$ is a monotonically increasing function and therefore as long as $F'(R)>0$, we obtain ${\tilde F}'(R)>0$. 
Because,
\begin{align}
\label{md2}
{\tilde F}'(R) \equiv \left( F'(R) + \frac{2{R_1}^2 F(R)}{\left( R - R_0 \right)^3} \right) \e^{ - \frac{{R_1}^2}{\left( R - R_0 \right)^2}}
\sim \frac{2{R_1}^2 F(R)}{\left( R - R_0 \right)^3}\e^{ - \frac{{R_1}^2}{\left( R- R_0 \right)^2}} \quad \mbox{when} \quad R\sim R_0 \, ,
\end{align}
when $R\sim R_0$, the potential $V$ in \eqref{JGRG23} behaves as,
\begin{align}
\label{JGRG23B}
V(A) = \frac{A}{{\tilde F}'(A)}  - \frac{\tilde F(A)}{{\tilde F}'(A)^2}
\sim \frac{A_0\left( A - R_0 \right)^3}{2{R_1}^2 F(R_0)}\e^{\frac{{R_1}^2}{\left( A - R_0 \right)^2}}
 - \frac{\left( A - R_0 \right)^6}{4 {R_1}^4 F(R_0)}\e^{\frac{{R_1}^2}{\left( A - R_0 \right)^2}}
\sim \frac{A_0\left( A - R_0  \right)^3}{2{R_1}^2 F(R_0)}\e^{\frac{{R_1}^2}{\left( A - R_0 \right)^2}} \, .
\end{align}
Because $\e^{\frac{{R_1}^2}{\left( A - R_0 \right)^2}}$ diverges very rapidly when $A\to R_0$, $V(A)$ also diverges positively and therefore the potential barrier appears at $A=R_0$, and therefore the Universe does not develop into the unphysical region. 
A possible quantum cosmological motivation for the prescription in Eq.~\eqref{md} follows from treating the boundary of the ghost-free sector as a limiting surface of the gravitational configuration space. 
Since crossing into $F_R<0$ reverses the sign of the graviton kinetic term, it is natural to consider quantum states whose support is suppressed near this boundary. 
In the Einstein frame, the relevant degree of freedom is the canonically normalized scalaron $\phi$ with potential $U(\phi)$, defined for the modified theory by
\begin{align}
\phi=-\frac{1}{\kappa}\sqrt{\frac{3}{2}}\ln\widetilde F_R \qquad U(\phi)=\frac{R\widetilde F_R-\widetilde F}{2\kappa^2\widetilde F_R^2} \, .
\end{align}
For a spatially flat FLRW minisuperspace with unit comoving volume, neglecting matter and factor ordering ambiguities, the Wheeler-DeWitt equation becomes
\begin{align}
\left[\frac{\kappa^2\hbar^2}{12\widetilde a}\frac{\partial^2}{\partial\widetilde a^2}-\frac{\hbar^2}{2\widetilde a^3}\frac{\partial^2}{\partial\phi^2}+\widetilde a^3U(\phi)\right]\Psi(\widetilde a,\phi)=0 \, ,
\end{align}
where $\widetilde a$ denotes the Einstein frame scale factor. Quantum cosmological investigations of future singularities have demonstrated that suitable boundary conditions and exponentially suppressed wave functions can exclude classically pathological configurations \cite{
Kiefer:2008sw, Kamenshchik:2007zj, Bouhmadi-Lopez:2009ggt, Alonso-Serrano:2018zpi, BorislavovVasilev:2021elf, Kiefer:2019bxk}. 
Motivated by this perspective, one may impose a DeWitt-type condition at a limiting curvature $R_0$ lying on the physical side of the anti-gravity transition, such that
\begin{align}
\lim_{R\to R_0^+}\Psi(\widetilde a,\phi(R))=0 \, .
\end{align}
Although this boundary does not correspond to a curvature singularity, the condition expresses the requirement that the quantum state have vanishing amplitude at the boundary of the admissible gravitational sector. 
In the semiclassical approximation, and for an adiabatically varying scale factor, a sufficiently large positive scalaron potential admits a decaying WKB branch whose leading behavior is
\begin{align}
\Psi(\widetilde a,\phi)\propto\exp\left[-\frac{\sqrt{2}\widetilde a^3}{\hbar}\int_{\phi_*}^{\phi}d\varphi\sqrt{U(\varphi)}\right] \, ,
\end{align}
where the expression applies asymptotically when the potential dominates the remaining terms in the scalaron equation. 
This suggests that a rapidly increasing positive potential can dynamically suppress configurations approaching the forbidden boundary. 
The exponential deformation introduced in Eq.~\eqref{md} provides an effective realization of this requirement and its multiplicative factor approaches unity away from $R_0$, while vanishing faster than any power as $R\to R_0$. It is easy to check that the leading behavior of the derivative of the modified gravitational function is
\begin{align}
\widetilde F_R\simeq\frac{2R_1^2F(R_0)}{(R-R_0)^3}\exp\left[-\frac{R_1^2}{(R-R_0)^2}\right] \, .
\end{align}
Consequently, the canonically normalized scalaron potential behaves as
\begin{align}
U(\phi(R))\simeq\frac{R_0(R-R_0)^3}{4\kappa^2R_1^2F(R_0)}\exp\left[\frac{R_1^2}{(R-R_0)^2}\right] \, .
\end{align}
For $R\to R_0^+$ with $R_0>0$ and $F(R_0)>0$, both $\phi$ and $U(\phi)$ diverge positively, giving us an exponentially suppressed WKB branch consistent with the proposed boundary condition. 
So, Eq.~\eqref{md} can be regarded as a phenomenological implementation of a quantum cosmological suppression principle rather than a uniquely derived quantum correction, providing additional motivation for introducing the potential barrier before the onset of the ghost graviton regime.

\section{Future Selection with Peculiar Velocities and Pairwise kSZ Probes} \label{sec8}

Having discussed the theoretical setup to avoid inconsistent physical futures, it is important to examine how observations can constrain the future evolution of the Universe in this regime. 
In particular, although the modified gravitational action in Eq.~\eqref{md} is constructed to preserve the cosmological expansion history preceding the potential barrier, its higher curvature derivatives can modify the dynamics of matter perturbations. 
Peculiar velocities and the pairwise kinematic Sunyaev-Zel'dovich (kSZ) effect provide a complementary means of probing these modifications, since they respond directly to the gravitational acceleration of nonrelativistic matter rather than solely to the background expansion. 
We hence investigate whether these observables can constrain the parameters controlling the curvature barrier and consequently select among theoretically admissible future cosmological trajectories. 
The relevant perturbative framework and its connection to large-scale velocities and kSZ observables have been developed extensively in the literature~\cite{
Song:2006ej, Hu:2007nk, 
Koyama:2015vza, Mueller:2014nsa, Bianchini:2015iaa}. 
We begin with the modified gravitational function introduced in Eq.~\eqref{md}, written as
\begin{align}
\widetilde F(R)=F(R)B(R) \, , \quad B(R)=\exp\left[-\frac{R_1^2}{(R-R_0)^2}\right] \, ,
\end{align}
where $R_0$ specifies the limiting curvature and $R_1$ determines the width of the modification. 
Defining $\Delta=R-R_0$, direct differentiation gives
\begin{align}
\widetilde F_R=B(R)\left[F_R+\frac{2R_1^2F}{\Delta^3}\right] \, ,
\end{align}
and
\begin{align}
\widetilde F_{RR}=B(R)\left[F_{RR}+\frac{4R_1^2F_R}{\Delta^3}+\left(\frac{4R_1^4}{\Delta^6}-\frac{6R_1^2}{\Delta^4}\right)F\right] \, .
\end{align}
The important observation is that the modification of the gravitational action and that of its curvature derivatives need not be comparable. 
Even when $R_1^2/\Delta^2\ll1$, the additional inverse powers of $\Delta$ can enhance the response of $\widetilde F_{RR}$ relative to that of $\widetilde F_R$. 
Consequently, a background expansion history close to that of the original model does not necessarily imply an equally small modification of its scalar perturbations.

For an explicit illustration of this derivative sensitivity, consider the logarithmic model in Eq.~\eqref{lnmodel} neglecting its inflationary $R^2/M^2$ contribution at late times and introduce $x=R/(2\Lambda)$, $x_0=R_0/(2\Lambda)$ and $r_1=R_1/(2\Lambda)$. 
We obtain the following
\begin{align}
F(R)\simeq2\Lambda\left[x-1+\frac{1}{1+\gamma\ln(\alpha x)}\right] \, .
\end{align}
Taking $\alpha=0.80$, $\gamma=1.483$, $x_0=1.6$ and $r_1=0.02$ and evaluating the derivatives at the reference curvature $x\simeq 2.214$, gives us $F_R\simeq 0.8039$, $\widetilde F_R\simeq 0.8091$,
$(2\Lambda)F_{RR}\simeq 0.2307$, $(2\Lambda)\widetilde F_{RR}\simeq 0.2065$.
Thus, the barrier changes the first derivative by approximately $0.65\%$, whereas the second derivative changes by approximately $-10.5\%$. 
The reference curvature corresponds to a flat $\Lambda$CDM proxy with $\Omega_{m0}=0.3$ and $\Omega_{\Lambda0}=0.7$ and is not obtained from a self-consistent solution of the modified Friedmann equations. 
This calculation illustrates the sensitivity to Eq.~\eqref{md}, but does not establish the observational viability of the chosen logarithmic model.

The connection with peculiar velocities follows from the scalar degree of freedom present in metric $F(R)$ gravity. 
Around a slowly varying background, the scalaron mass is approximately
\begin{align}
m_s^2=\frac{1}{3}\left(\frac{\widetilde F_R}{\widetilde F_{RR}}-R\right) \, .
\end{align}
For a stable branch with $m_s^2>0$, the scalaron mediates an additional gravitational interaction characterized by an effective Compton wavelength $\lambda_C=m_s^{-1}$. 
Since the curvature barrier modifies both $\widetilde F_R$ and $\widetilde F_{RR}$, it can change the scale on which this interaction becomes important. 
The sensitivity of the scalaron mass to $\widetilde F_{RR}$ is particularly useful because this quantity need not be strongly constrained by measurements of $H(z)$ alone. 
In the subhorizon quasi-static regime, the modified Poisson equation for nonrelativistic matter can be expressed as
\begin{align}
k^2\Psi=-4\pi Ga^2\mu(k,a)\rho_m\delta_m \, ,
\end{align}
where the effective gravitational strength is
\begin{align}
\mu(k,a)=\frac{1}{\widetilde F_R}\frac{1+4Q(k,a)}{1+3Q(k,a)} \qquad Q(k,a)=\frac{k^2}{a^2}\frac{\widetilde F_{RR}}{\widetilde F_R} \, .
\end{align}
Here $k$ is the comoving wavenumber and $\delta_m$ is the matter density contrast, and the scale dependence arises because different Fourier modes respond differently to the scalaron-mediated force.  
Therefore, two cosmological models with nearly identical background expansion histories may predict different gravitational accelerations and consequently different coherent matter velocities. 
Going forward, combining the linear continuity and Euler equations with the modified Poisson equation gives
\begin{align}
\ddot{\delta}_m+2H\dot{\delta}_m-4\pi G\rho_m\mu(k,a)\delta_m=0 \, .
\end{align}
Introducing the scale-dependent growth factor and growth rate through
\begin{align}
\delta_m(k,a)=D(k,a)\delta_m(k,a_i) \qquad f(k,a)=\frac{d\ln D(k,a)}{d\ln a} \, , 
\end{align}
we obtain, equivalently
\begin{align}
\frac{d^2D}{d(\ln a)^2}+\left(2+\frac{d\ln H}{d\ln a}\right)\frac{dD}{d\ln a}-\frac{3}{2}\Omega_m(a)\mu(k,a)D=0 \, .
\end{align}
The consequence is that even a modest change in $\mu(k, a)$ can accumulate over cosmological time into a measurable modification of the growth rate. 
Importantly, this change can remain scale-dependent, unlike a simple rescaling of the primordial matter fluctuation amplitude. 
This allows velocity statistics measured over different separations and redshifts to probe the curvature dependence introduced by the barrier. 
For irrotational linear perturbations, the Fourier space peculiar velocity field follows from the continuity equation as
\begin{align}
\mathbf v(\mathbf k,a)=iaHf(k,a)\delta_m(\mathbf k,a)\frac{\mathbf k}{k^2} \, .
\end{align}
Its longitudinal velocity power spectrum is consequently,
\begin{align}
P_{vv}(k,a)=\frac{a^2H^2f^2(k,a)}{k^2}P_{\delta\delta}(k,a) \, ,
\end{align}
where $P_{\delta\delta}$ is the linear matter power spectrum. 
The velocity spectrum thus receives modifications from both the instantaneous growth rate and the accumulated density growth. 
This provides a direct observational link between the curvature barrier and coherent cosmic motions, which can be investigated independently of the background distance-redshift relation.

The corresponding mean pairwise radial velocity of matter at comoving separation $r$ is, at leading order in linear perturbations,
\begin{align}
v_{12}(r,a)=-\frac{aH}{\pi^2[1+\xi_m(r,a)]}\int_0^\infty dk\,k\,f(k,a)P_{\delta\delta}(k,a)j_1(kr) \, ,
\end{align}
where $\xi_m$ is the matter two-point correlation function and $j_1$ is the spherical Bessel function of the first order. 
For a scale-independent growth rate, this reduces to the familiar expression 
\begin{align}
v_{12}(r,a)=-\frac{2}{3}aHf(a)r\frac{\overline{\xi}_m(r,a)}{1+\xi_m(r,a)} \, ,
\end{align}
with $\overline{\xi}_m(r,a)=3r^{-3}\int_0^r ds\,s^2\xi_m(s,a)$. 
The scale-dependent expression is more relevant for the present construction because the scalaron force modifies different wavelengths differently. 
For a linearly biased galaxy sample, the numerator acquires a factor of the effective bias while the denominator involves the tracer correlation function, with additional corrections required on nonlinear scales. 
The pairwise kSZ effect offers a direct observational realization of this velocity statistic, and the temperature perturbation produced by Thomson scattering from moving electrons is, 
\begin{align}
\frac{\Delta T_\mathrm{kSZ}}{T_\mathrm{CMB}}=-\frac{\sigma_T}{c}\int d\ell\,n_e(\ell)v_\parallel(\ell) \, ,
\end{align}
where $\sigma_T$ is the Thomson cross section, $n_e$ is the physical electron density and $v_\parallel$ is its line of sight peculiar velocity. 
For a localized object with effective optical depth $\tau_e$, this approximately becomes
\begin{align}
\Delta T_\mathrm{kSZ}\simeq-T_\mathrm{CMB}\tau_e\frac{v_\parallel}{c} \, .
\end{align}
Measurements of the pairwise kSZ signal statistically isolate the coherent relative motions of galaxy and cluster populations. 
The observed pairwise temperature amplitude is approximately proportional to the mean pairwise velocity, after accounting for the optical depth distribution, filtering, and geometric projection. 
The initial detection of this effect and subsequent simulation studies provide the observational and astrophysical basis for such an interpretation~\cite{Hand:2012ui, Flender:2015btu
}.

To estimate the prospective magnitude of the effect, consider a schematic stable and nearly GR-normalized branch with $\widetilde F_R\simeq1$. 
In the regime where the curvature contribution to the scalaron mass is subdominant, the effective gravitational strength takes the Yukawa-like form $\mu(k,a)\simeq1+\frac{1}{3}\frac{k^2}{k^2+a^2m_s^2}$. 
For an illustrative present-epoch comoving scalaron range $\lambda_C=10\,h^{-1}\mathrm{Mpc}$, corresponding to $k_C=am_s\simeq0.1\,h\,\mathrm{Mpc}^{-1}$ we find $\left\{\mu(0.05),\mu(0.10),\mu(0.20)\right\}\simeq\left\{1.067,1.167,1.267\right\}$ where the wavenumbers are expressed in $h\,\mathrm{Mpc}^{-1}$. 
The corresponding enhancements of the effective gravitational force are approximately $6.7\%$, $16.7\%$ and $26.7\%$. 
These values illustrate the scale-dependent response permitted by the scalaron sector and are not predictions from the logarithmic parameter choice considered above. 
In particular, obtaining this force range from Eq.~\eqref{md} requires a separate stable and observationally viable background solution.

A simple estimate of the resulting velocity enhancement can be obtained by considering a matter-dominated approximation with an effectively constant $\mu$ over a short interval. Writing $\delta_m\propto a^p$, the growth equation gives
\begin{align}
p=\frac{\sqrt{1+24\mu}-1}{4} \, .
\end{align}
At $k=0.1\,h\,\mathrm{Mpc}^{-1}$, the illustrative value $\mu\simeq1.167$ gives $p\simeq1.096$, compared with $p=1$ for general relativity in the same matter-dominated approximation. 
If the perturbations are normalized equally at $z_i=0.5$, the accumulated growth enhancement by $z=0$ is 
\begin{align}
\frac{D_\mathrm{MG}(0)}{D_\mathrm{GR}(0)}\simeq1.5^{p-1}\simeq1.040 \, .
\end{align}
The corresponding velocity amplitude and velocity power ratios are then
\begin{align}
\frac{v_\mathrm{MG}}{v_\mathrm{GR}}\simeq p\,1.5^{p-1}\simeq1.140 \, , 
\end{align}
\begin{align}
\frac{P_{vv}^\mathrm{MG}}{P_{vv}^\mathrm{GR}}\simeq p^2\,1.5^{2(p-1)}\simeq1.30 \, .
\end{align}
Thus, a $16.7\%$ gravitational force enhancement can generate approximately $14\%$ larger coherent velocities and $30\%$ larger velocity power in this deliberately simplified calculation. 
The numerical estimate assumes a matter-dominated expansion, a fixed effective coupling, and a narrow range of Fourier modes. 
A realistic late-time prediction must instead integrate the full scale- and redshift-dependent growth equations with cosmic acceleration and any relevant screening effects included. 
For the kSZ signal, consider representative values $\tau_e=10^{-3}$ and $|v_\parallel|=300\,\mathrm{km\,s^{-1}}$. With $T_\mathrm{CMB}=2.7255\,\mathrm{K}$, the characteristic single object temperature amplitude is
\begin{align}
|\Delta T_\mathrm{kSZ}|\simeq2.73\,\mu\mathrm{K} \, .
\end{align}
If a comparable $14\%$ velocity enhancement were present while the optical depth remained fixed, the corresponding temperature shift would be
\begin{align}
|\delta T_\mathrm{kSZ}|\simeq0.14\times2.73\,\mu\mathrm{K}\simeq0.38\,\mu\mathrm{K} \, .
\end{align}
This is an illustrative single-object amplitude difference, not the predicted amplitude of a pairwise kSZ estimator. 
The latter involves the full separation-dependent velocity correlation, projection factors, and optical depth weighting. 
Nevertheless, it demonstrates that relatively small changes in gravitational dynamics can translate into potentially measurable statistical differences when large samples of galaxies and halos are combined.

The observational prospects are particularly relevant in view of recent measurements, as the combination of DESI DR1 luminous red galaxies with ACT DR6 and Planck temperature maps has yielded a $9.3\sigma$ detection of the pairwise kSZ effect using approximately $9.13\times10^5$ galaxies \cite{Gong:2025ffw}. 
Independent DESI DR1 analyses have also obtained the low-redshift peculiar velocity growth constraint $f\sigma_8(z_\mathrm{eff}=0.07)=0.450\pm0.055$ \cite{Lai:2025xkf
}, while measurements using DESI bright galaxies and ACT have further demonstrated the applicability of pairwise kSZ analyses to large galaxy samples \cite{
Hadzhiyska:2025egz}. 
These measurements establish the observational feasibility of the proposed test, although the statistical significance of a kSZ detection should not be identified with an equally precise determination of peculiar velocities because of optical depth and astrophysical degeneracies. 
A suitable observable for isolating the barrier-induced signal is the ratio of the predicted pairwise velocities,
\begin{align}
\mathcal V(r,z;\Theta)=\frac{v_{12}^\mathrm{barrier}(r,z;\Theta)}{v_{12}^\mathrm{reference}(r,z)}
\end{align}
where $\Theta$ denotes the gravitational model parameters, including $R_0$ and $R_1$. 
The scale and redshift dependence of $\mathcal V$ can also help distinguish the effects of a changing scalaron force range from an overall normalization of matter fluctuations. 
In practice, the prediction must also account for tracer bias, optical depth, baryonic physics and the survey selection function, which can otherwise imitate changes in the pairwise signal. 
We can hence formulate future selection as an inference problem and let $D$ denote the combined peculiar velocity and pairwise kSZ measurements, and let $\mathcal I_\mathrm{safe}(\Theta)$ equal unity for models whose cosmological evolution remains on a dynamically consistent ghost-free branch and vanish otherwise. 
The probability assigned to such future trajectories is
\begin{align}
P(\mathrm{safe}\mid D)=\frac{\int d\Theta\,\mathcal I_\mathrm{safe}(\Theta)\mathcal L_\mathrm{PV}(D\mid\Theta) 
\mathcal L_\mathrm{kSZ}(D\mid\Theta)\pi(\Theta)}{\int d\Theta\,\mathcal L_\mathrm{PV}(D\mid\Theta)\mathcal L_\mathrm{kSZ}(D\mid\Theta)\pi(\Theta)}
\end{align}
where $\pi(\Theta)$ is the model prior, and the likelihoods should be replaced by a joint likelihood when the datasets are correlated. 
The indicator requires solving the future background dynamics and testing the relevant scalar and tensor stability conditions, as it cannot be inferred solely from the existence of a divergent scalaron potential. 
This construction makes it very clear how present observations can preferentially select different future cosmological histories within a specified family of modified gravity theories.

\section{Summary and Discussions}

In this paper, we found that in some realistic models~\cite{Odintsov:2026fyc}, the anti-gravity or ghost graviton often appears in the future for these models. 
Although the problem could be avoided by further changing the parameters and fine-tuning the initial conditions, we propose to introduce a potential barrier of the scalar field for the small scalar curvature region, which prevents the Universe from evolving into the anti-gravity or ghost graviton region. 
This mechanism does not need the fine-tuning of the initial conditions. 
We may also include the effects coming from matter, but if we consider dark matter, for example. 
Because the energy density of the dark matter, however, behaves as $a^{-3}$,  the contributions become very small in the future, and we may neglect the effects.

As we mention in the introduction, we assumed that the evolution of the Universe after the (inverse) phantom crossing is given by the asymptotic de~Sitter spacetime or the decelerating Universe. 
There are many other possibilities; that is, the expanding Universe might begin to contract, and it may go to the Big Crunch or continue to contract forever. 
Or the Universe might continue to expand and go to the asymptotic de~Sitter spacetime, including the pseudo-rip; it could continue to expand accelerated but not go to the asymptotic de~Sitter spacetime, or it could go to a singularity in the finite future. 
There is also a possibility of the oscillating Universe~\cite{Nojiri:2025low} or other more complex behaviors in the future. We have also remarked in the section above, how future selection for the cosmology we have described above could be probed using kSZ and peculiar velocity effects.

A similar mechanism was used in \cite{Bamba:2011sm} in order to avoid the future singularity~\cite{Caldwell:1999ew, Caldwell:2003vq, Barrow:2004xh, Nojiri:2004pf, Nojiri:2005sx} (see \cite{deHaro:2023lbq} for review) of the Universe, where the potential barrier appears for large scalar curvature regions.
Suppose we consider a model in which, at points where the scalar curvature is very large or very small, the potential of the scalar field becomes very large or diverges. 
In that case, classically, the scalar curvature cannot become very large or very small. 
In path integrals of quantum theory, this potential suppresses contributions from regions where the scalar curvature is very large or very small; consequently, it appears to act as a cut-off that suppresses ultraviolet and infrared divergences. 
This mechanism may help to define a quantum theory of gravity. 
We should note that perturbative analysis is difficult, but if we can define a theory free of divergences, we believe the quantum corrections would also be finite. 
Of course, this mechanism does not always prevent the appearance of the curvature singularity.
For example, the scalar curvature vanishes in the Schwarzschild spacetime but, as is very well known, the Kretschmann scalar diverges at the center of the Schwarzschild spacetime, and there is a curvature singularity there. 
Furthermore, the divergence coming from the tensor mode could not be suppressed at least perturbatively. 
We should also note the fact that a large $F(R)$ does not correspond to a large potential of the scalar field. 
In path integrals, the measure over which the integral is taken may change as a result of the field-dependent Weyl rescaling.

We believe the relationship between suppressing high- or low-curvature configurations and suppressing ultraviolet or infrared divergences in quantum theory needs to be examined in somewhat greater detail. 
Anyway, as this line of thinking approaches the definition of quantum gravity from a perspective distinct from perturbative renormalization, we believe it is worth continuing to explore.

\section*{Acknowledgments}

SO acknowledges support from the Arrakihs mission through grant
number PID2025-168567OB-C62 of the Spanish Ministerio de Ciencia,
Innovaci\'{o}n y Universidades.

\bibliographystyle{apsrev4-1}
\bibliography{References4-2}

@article{Nojiri:2005vv,
    author = "Nojiri, Shin'ichi and Odintsov, Sergei D. and Sasaki, Misao",
    title = "{Gauss-Bonnet dark energy}",
    eprint = "hep-th/0504052",
    archivePrefix = "arXiv",
    reportNumber = "YITP-05-14",
    doi = "10.1103/PhysRevD.71.123509",
    journal = "Phys. Rev. D",
    volume = "71",
    pages = "123509",
    year = "2005"
}

@article{Nojiri:2005pu,
    author = "Nojiri, Shin'ichi and Odintsov, Sergei D.",
    title = "{Unifying phantom inflation with late-time acceleration: Scalar phantom-non-phantom transition model and generalized holographic dark energy}",
    eprint = "hep-th/0506212",
    archivePrefix = "arXiv",
    doi = "10.1007/s10714-006-0301-6",
    journal = "Gen. Rel. Grav.",
    volume = "38",
    pages = "1285--1304",
    year = "2006"
}

@article{Capozziello:2005tf,
    author = "Capozziello, S. and Nojiri, S. and Odintsov, S. D.",
    title = "{Unified phantom cosmology: Inflation, dark energy and dark matter under the same standard}",
    eprint = "hep-th/0507182",
    archivePrefix = "arXiv",
    doi = "10.1016/j.physletb.2005.11.012",
    journal = "Phys. Lett. B",
    volume = "632",
    pages = "597--604",
    year = "2006"
}

@article{Nojiri:2005jg,
    author = "Nojiri, Shin'ichi and Odintsov, Sergei D.",
    title = "{Modified Gauss-Bonnet theory as gravitational alternative for dark energy}",
    eprint = "hep-th/0508049",
    archivePrefix = "arXiv",
    doi = "10.1016/j.physletb.2005.10.010",
    journal = "Phys. Lett. B",
    volume = "631",
    pages = "1--6",
    year = "2005"
}

@article{Nojiri:2009kx,
    author = "Nojiri, Shin'ichi and Odintsov, Sergei D. and Saez-Gomez, Diego",
    title = "{Cosmological reconstruction of realistic modified F(R) gravities}",
    eprint = "0908.1269",
    archivePrefix = "arXiv",
    primaryClass = "hep-th",
    doi = "10.1016/j.physletb.2009.09.045",
    journal = "Phys. Lett. B",
    volume = "681",
    pages = "74--80",
    year = "2009"
}

@article{Nojiri:2010wj,
    author = "Nojiri, Shin'ichi and Odintsov, Sergei D.",
    title = "{Unified cosmic history in modified gravity: from F(R) theory to Lorentz non-invariant models}",
    eprint = "1011.0544",
    archivePrefix = "arXiv",
    primaryClass = "gr-qc",
    doi = "10.1016/j.physrep.2011.04.001",
    journal = "Phys. Rept.",
    volume = "505",
    pages = "59--144",
    year = "2011"
}

@article{Capozziello:2011et,
    author = "Capozziello, Salvatore and De Laurentis, Mariafelicia",
    title = "{Extended Theories of Gravity}",
    eprint = "1108.6266",
    archivePrefix = "arXiv",
    primaryClass = "gr-qc",
    doi = "10.1016/j.physrep.2011.09.003",
    journal = "Phys. Rept.",
    volume = "509",
    pages = "167--321",
    year = "2011"
}

@article{Nojiri:2017ncd,
    author = "Nojiri, S. and Odintsov, S. D. and Oikonomou, V. K.",
    title = "{Modified Gravity Theories on a Nutshell: Inflation, Bounce and Late-time Evolution}",
    eprint = "1705.11098",
    archivePrefix = "arXiv",
    primaryClass = "gr-qc",
    reportNumber = "PHYS.REPT.-692-(2017)-1-104, Phys.Rept. 692 (2017) 1-104",
    doi = "10.1016/j.physrep.2017.06.001",
    journal = "Phys. Rept.",
    volume = "692",
    pages = "1--104",
    year = "2017"
}

@article{Nojiri:2003ft,
    author = "Nojiri, Shin'ichi and Odintsov, Sergei D.",
    title = "{Modified gravity with negative and positive powers of the curvature: Unification of the inflation and of the cosmic acceleration}",
    eprint = "hep-th/0307288",
    archivePrefix = "arXiv",
    doi = "10.1103/PhysRevD.68.123512",
    journal = "Phys. Rev. D",
    volume = "68",
    pages = "123512",
    year = "2003"
}

@article{Starobinsky:1980te,
    author = "Starobinsky, Alexei A.",
    editor = "Khalatnikov, I. M. and Mineev, V. P.",
    title = "{A New Type of Isotropic Cosmological Models Without Singularity}",
    doi = "10.1016/0370-2693(80)90670-X",
    journal = "Phys. Lett. B",
    volume = "91",
    pages = "99--102",
    year = "1980"
}

@article{Nojiri:2007cq,
    author = "Nojiri, Shin'ichi and Odintsov, Sergei D.",
    title = "{Modified f(R) gravity unifying R**m inflation with Lambda CDM epoch}",
    eprint = "0710.1738",
    archivePrefix = "arXiv",
    primaryClass = "hep-th",
    reportNumber = "YITP-07-58",
    doi = "10.1103/PhysRevD.77.026007",
    journal = "Phys. Rev. D",
    volume = "77",
    pages = "026007",
    year = "2008"
}

@article{DESI:2024mwx,
    author = "Adame, A. G. and others",
    collaboration = "DESI",
    title = "{DESI 2024 VI: cosmological constraints from the measurements of baryon acoustic oscillations}",
    eprint = "2404.03002",
    archivePrefix = "arXiv",
    primaryClass = "astro-ph.CO",
    reportNumber = "FERMILAB-PUB-24-0154-PPD",
    doi = "10.1088/1475-7516/2025/02/021",
    journal = "JCAP",
    volume = "02",
    pages = "021",
    year = "2025"
}

@article{DESI:2025zgx,
    author = "Abdul Karim, M. and others",
    collaboration = "DESI",
    title = "{DESI DR2 results. II. Measurements of baryon acoustic oscillations and cosmological constraints}",
    eprint = "2503.14738",
    archivePrefix = "arXiv",
    primaryClass = "astro-ph.CO",
    reportNumber = "FERMILAB-PUB-25-0169-PPD",
    doi = "10.1103/tr6y-kpc6",
    journal = "Phys. Rev. D",
    volume = "112",
    number = "8",
    pages = "083515",
    year = "2025"
}

@article{Oikonomou:2025qub,
    author = "Oikonomou, V. K.",
    title = "{Towards model agnostic F(R) gravity inflation}",
    eprint = "2504.00915",
    archivePrefix = "arXiv",
    primaryClass = "gr-qc",
    doi = "10.1088/1475-7516/2025/10/113",
    journal = "JCAP",
    volume = "10",
    pages = "113",
    year = "2025"
}

@article{Nojiri:2025low,
    author = "Nojiri, Shin'ichi and Odintsov, S. D. and Oikonomou, V. K.",
    title = "{Phantom crossing and oscillating dark energy with F(R) gravity}",
    eprint = "2506.21010",
    archivePrefix = "arXiv",
    primaryClass = "gr-qc",
    reportNumber = "KEK-TH-2734, KEK-Cosmo-0383",
    doi = "10.1103/16yg-966k",
    journal = "Phys. Rev. D",
    volume = "112",
    number = "10",
    pages = "104035",
    year = "2025"
}

@book{Faraoni:2010pgm,
    author = "Faraoni, Valerio and Capozziello, Salvatore",
    title = "{Beyond Einstein Gravity}: {A Survey of Gravitational Theories for Cosmology and Astrophysics}",
    doi = "10.1007/978-94-007-0165-6",
    isbn = "978-94-007-0164-9, 978-94-007-0165-6",
    publisher = "Springer",
    address = "Dordrecht",
    year = "2011"
}

@article{Nojiri:2006ri,
    author = "Nojiri, Shin'ichi and Odintsov, Sergei D.",
    editor = "Borowiec, Andrzej",
    title = "{Introduction to modified gravity and gravitational alternative for dark energy}",
    eprint = "hep-th/0601213",
    archivePrefix = "arXiv",
    reportNumber = "KARP-2006-06",
    doi = "10.1142/S0219887807001928",
    journal = "eConf",
    volume = "C0602061",
    pages = "06",
    year = "2006"
}

@article{Odintsov:2024woi,
    author = "Odintsov, Sergei D. and S{\'a}ez-Chill{\'o}n G{\'o}mez, Diego and Sharov, German S.",
    title = "{Modified gravity/dynamical dark energy vs $\Lambda $CDM: is the game over?}",
    eprint = "2412.09409",
    archivePrefix = "arXiv",
    primaryClass = "gr-qc",
    doi = "10.1140/epjc/s10052-025-14013-3",
    journal = "Eur. Phys. J. C",
    volume = "85",
    number = "3",
    pages = "298",
    year = "2025"
}

@article{Dai:2020rfo,
    author = "Dai, Wei-Ming and Ma, Yin-Zhe and He, Hong-Jian",
    title = "{Reconciling Hubble Constant Discrepancy from Holographic Dark Energy}",
    eprint = "2003.03602",
    archivePrefix = "arXiv",
    primaryClass = "astro-ph.CO",
    doi = "10.1103/PhysRevD.102.121302",
    journal = "Phys. Rev. D",
    volume = "102",
    pages = "121302",
    year = "2020"
}

@article{He:2020zns,
    author = "He, Hong-Jian and Ma, Yin-Zhe and Zheng, Jiaming",
    title = "{Resolving Hubble Tension by Self-Interacting Neutrinos with Dirac Seesaw}",
    eprint = "2003.12057",
    archivePrefix = "arXiv",
    primaryClass = "hep-ph",
    doi = "10.1088/1475-7516/2020/11/003",
    journal = "JCAP",
    volume = "11",
    pages = "003",
    year = "2020"
}

@article{Nakai:2020oit,
    author = "Nakai, Yuichiro and Suzuki, Motoo and Takahashi, Fuminobu and Yamada, Masaki",
    title = "{Gravitational Waves and Dark Radiation from Dark Phase Transition: Connecting NANOGrav Pulsar Timing Data and Hubble Tension}",
    eprint = "2009.09754",
    archivePrefix = "arXiv",
    primaryClass = "astro-ph.CO",
    reportNumber = "TU-1109; IPMU20-0100",
    doi = "10.1016/j.physletb.2021.136238",
    journal = "Phys. Lett. B",
    volume = "816",
    pages = "136238",
    year = "2021"
}

@article{DiValentino:2020naf,
    author = "Di Valentino, Eleonora and Mukherjee, Ankan and Sen, Anjan A.",
    title = "{Dark Energy with Phantom Crossing and the $H_0$ Tension}",
    eprint = "2005.12587",
    archivePrefix = "arXiv",
    primaryClass = "astro-ph.CO",
    reportNumber = "IPPP/20/89",
    doi = "10.3390/e23040404",
    journal = "Entropy",
    volume = "23",
    number = "4",
    pages = "404",
    year = "2021"
}

@article{Agrawal:2019dlm,
    author = "Agrawal, Prateek and Obied, Georges and Vafa, Cumrun",
    title = "{$H_0$ tension, swampland conjectures, and the epoch of fading dark matter}",
    eprint = "1906.08261",
    archivePrefix = "arXiv",
    primaryClass = "astro-ph.CO",
    doi = "10.1103/PhysRevD.103.043523",
    journal = "Phys. Rev. D",
    volume = "103",
    number = "4",
    pages = "043523",
    year = "2021"
}

@article{Ye:2020btb,
    author = "Ye, Gen and Piao, Yun-Song",
    title = "{Is the Hubble tension a hint of AdS phase around recombination?}",
    eprint = "2001.02451",
    archivePrefix = "arXiv",
    primaryClass = "astro-ph.CO",
    doi = "10.1103/PhysRevD.101.083507",
    journal = "Phys. Rev. D",
    volume = "101",
    number = "8",
    pages = "083507",
    year = "2020"
}

@article{Vagnozzi:2021tjv,
    author = "Vagnozzi, Sunny and Pacucci, Fabio and Loeb, Abraham",
    title = "{Implications for the Hubble tension from the ages of the oldest astrophysical objects}",
    eprint = "2105.10421",
    archivePrefix = "arXiv",
    primaryClass = "astro-ph.CO",
    doi = "10.1016/j.jheap.2022.07.004",
    journal = "JHEAp",
    volume = "36",
    pages = "27--35",
    year = "2022"
}

@article{Desmond:2019ygn,
    author = "Desmond, Harry and Jain, Bhuvnesh and Sakstein, Jeremy",
    title = "{Local resolution of the Hubble tension: The impact of screened fifth forces on the cosmic distance ladder}",
    eprint = "1907.03778",
    archivePrefix = "arXiv",
    primaryClass = "astro-ph.CO",
    doi = "10.1103/PhysRevD.100.043537",
    journal = "Phys. Rev. D",
    volume = "100",
    number = "4",
    pages = "043537",
    year = "2019",
    note = "[Erratum: Phys.Rev.D 101, 069904 (2020), Erratum: Phys.Rev.D 101, 129901 (2020)]"
}

@article{Hogas:2023pjz,
    author = {H{\"o}g{\r{a}}s, Marcus and M{\"o}rtsell, Edvard},
    title = "{Hubble tension and fifth forces}",
    eprint = "2309.01744",
    archivePrefix = "arXiv",
    primaryClass = "astro-ph.CO",
    doi = "10.1103/PhysRevD.108.124050",
    journal = "Phys. Rev. D",
    volume = "108",
    number = "12",
    pages = "124050",
    year = "2023"
}

@article{OColgain:2018czj,
    author = "{\'O} Colg{\'a}in, Eoin and van Putten, Maurice H. P. M. and Yavartanoo, Hossein",
    title = "{de Sitter Swampland, $H_0$ tension {\textbackslash}{\&} observation}",
    eprint = "1807.07451",
    archivePrefix = "arXiv",
    primaryClass = "hep-th",
    doi = "10.1016/j.physletb.2019.04.032",
    journal = "Phys. Lett. B",
    volume = "793",
    pages = "126--129",
    year = "2019"
}

@article{Vagnozzi:2019ezj,
    author = "Vagnozzi, Sunny",
    title = "{New physics in light of the $H_0$ tension: An alternative view}",
    eprint = "1907.07569",
    archivePrefix = "arXiv",
    primaryClass = "astro-ph.CO",
    doi = "10.1103/PhysRevD.102.023518",
    journal = "Phys. Rev. D",
    volume = "102",
    number = "2",
    pages = "023518",
    year = "2020"
}

@article{Krishnan:2020obg,
    author = "Krishnan, C. and Colg{\'a}in, Eoin {\'O}. and Ruchika and Sen, Anjan A. and Sheikh-Jabbari, M. M. and Yang, Tao",
    title = "{Is there an early Universe solution to Hubble tension?}",
    eprint = "2002.06044",
    archivePrefix = "arXiv",
    primaryClass = "astro-ph.CO",
    doi = "10.1103/PhysRevD.102.103525",
    journal = "Phys. Rev. D",
    volume = "102",
    number = "10",
    pages = "103525",
    year = "2020"
}

@article{Colgain:2019joh,
    author = "Colg{\'a}in, Eoin {\'O}. and Yavartanoo, Hossein",
    title = "{Testing the Swampland: $H_0$ tension}",
    eprint = "1905.02555",
    archivePrefix = "arXiv",
    primaryClass = "astro-ph.CO",
    doi = "10.1016/j.physletb.2019.134907",
    journal = "Phys. Lett. B",
    volume = "797",
    pages = "134907",
    year = "2019"
}

@article{Vagnozzi:2021gjh,
    author = "Vagnozzi, Sunny",
    title = "{Consistency tests of {\ensuremath{\Lambda}}CDM from the early integrated Sachs-Wolfe effect: Implications for early-time new physics and the Hubble tension}",
    eprint = "2105.10425",
    archivePrefix = "arXiv",
    primaryClass = "astro-ph.CO",
    doi = "10.1103/PhysRevD.104.063524",
    journal = "Phys. Rev. D",
    volume = "104",
    number = "6",
    pages = "063524",
    year = "2021"
}

@article{Lee:2022cyh,
    author = "Lee, Bum-Hoon and Lee, Wonwoo and Colg{\'a}in, Eoin {\'O}. and Sheikh-Jabbari, M. M. and Thakur, Somyadip",
    title = "{Is local H $_{0}$ at odds with dark energy EFT?}",
    eprint = "2202.03906",
    archivePrefix = "arXiv",
    primaryClass = "astro-ph.CO",
    doi = "10.1088/1475-7516/2022/04/004",
    journal = "JCAP",
    volume = "04",
    number = "04",
    pages = "004",
    year = "2022"
}

@article{Krishnan:2021dyb,
    author = "Krishnan, Chethan and Mohayaee, Roya and Colg{\'a}in, Eoin {\'O}. and Sheikh-Jabbari, M. M. and Yin, Lu",
    title = "{Does Hubble tension signal a breakdown in FLRW cosmology?}",
    eprint = "2105.09790",
    archivePrefix = "arXiv",
    primaryClass = "astro-ph.CO",
    doi = "10.1088/1361-6382/ac1a81",
    journal = "Class. Quant. Grav.",
    volume = "38",
    number = "18",
    pages = "184001",
    year = "2021"
}

@article{Ye:2021iwa,
    author = "Ye, Gen and Zhang, Jun and Piao, Yun-Song",
    title = "{Alleviating both H0 and S8 tensions: Early dark energy lifts the CMB-lockdown on ultralight axion}",
    eprint = "2107.13391",
    archivePrefix = "arXiv",
    primaryClass = "astro-ph.CO",
    doi = "10.1016/j.physletb.2023.137770",
    journal = "Phys. Lett. B",
    volume = "839",
    pages = "137770",
    year = "2023"
}

@article{Ye:2022afu,
    author = "Ye, Gen and Piao, Yun-Song",
    title = "{Improved constraints on primordial gravitational waves in light of the H0 tension and BICEP/Keck data}",
    eprint = "2202.10055",
    archivePrefix = "arXiv",
    primaryClass = "astro-ph.CO",
    doi = "10.1103/PhysRevD.106.043536",
    journal = "Phys. Rev. D",
    volume = "106",
    number = "4",
    pages = "043536",
    year = "2022"
}

@article{Verde:2019ivm,
    author = "Verde, L. and Treu, T. and Riess, A. G.",
    title = "{Tensions between the Early and the Late Universe}",
    eprint = "1907.10625",
    archivePrefix = "arXiv",
    primaryClass = "astro-ph.CO",
    doi = "10.1038/s41550-019-0902-0",
    journal = "Nature Astron.",
    volume = "3",
    pages = "891",
    year = "2019"
}

@article{Menci:2024rbq,
    author = "Menci, Nicola and Adil, Shahnawaz A. and Mukhopadhyay, Upala and Sen, Anjan A. and Vagnozzi, Sunny",
    title = "{Negative cosmological constant in the dark energy sector: tests from JWST photometric and spectroscopic observations of high-redshift galaxies}",
    eprint = "2401.12659",
    archivePrefix = "arXiv",
    primaryClass = "astro-ph.CO",
    doi = "10.1088/1475-7516/2024/07/072",
    journal = "JCAP",
    volume = "07",
    pages = "072",
    year = "2024"
}

@article{Adil:2023ara,
    author = "Adil, Shahnawaz A. and Mukhopadhyay, Upala and Sen, Anjan A. and Vagnozzi, Sunny",
    title = "{Dark energy in light of the early JWST observations: case for a negative cosmological constant?}",
    eprint = "2307.12763",
    archivePrefix = "arXiv",
    primaryClass = "astro-ph.CO",
    doi = "10.1088/1475-7516/2023/10/072",
    journal = "JCAP",
    volume = "10",
    pages = "072",
    year = "2023"
}

@article{Reeves:2022aoi,
    author = "Reeves, Alexander and Herold, Laura and Vagnozzi, Sunny and Sherwin, Blake D. and Ferreira, Elisa G. M.",
    title = "{Restoring cosmological concordance with early dark energy and massive neutrinos?}",
    eprint = "2207.01501",
    archivePrefix = "arXiv",
    primaryClass = "astro-ph.CO",
    doi = "10.1093/mnras/stad317",
    journal = "Mon. Not. Roy. Astron. Soc.",
    volume = "520",
    number = "3",
    pages = "3688--3695",
    year = "2023"
}

@article{Ferlito:2022mok,
    author = "Ferlito, Fulvio and Vagnozzi, Sunny and Mota, David F. and Baldi, Marco",
    title = "{Cosmological direct detection of dark energy: Non-linear structure formation signatures of dark energy scattering with visible matter}",
    eprint = "2201.04528",
    archivePrefix = "arXiv",
    primaryClass = "astro-ph.CO",
    doi = "10.1093/mnras/stac649",
    journal = "Mon. Not. Roy. Astron. Soc.",
    volume = "512",
    number = "2",
    pages = "1885--1905",
    year = "2022"
}

@article{Vagnozzi:2021quy,
    author = "Vagnozzi, Sunny and Visinelli, Luca and Brax, Philippe and Davis, Anne-Christine and Sakstein, Jeremy",
    title = "{Direct detection of dark energy: The XENON1T excess and future prospects}",
    eprint = "2103.15834",
    archivePrefix = "arXiv",
    primaryClass = "hep-ph",
    doi = "10.1103/PhysRevD.104.063023",
    journal = "Phys. Rev. D",
    volume = "104",
    number = "6",
    pages = "063023",
    year = "2021"
}

@article{DiValentino:2020evt,
    author = "Di Valentino, Eleonora and Gariazzo, Stefano and Mena, Olga and Vagnozzi, Sunny",
    title = "{Soundness of Dark Energy properties}",
    eprint = "2005.02062",
    archivePrefix = "arXiv",
    primaryClass = "astro-ph.CO",
    doi = "10.1088/1475-7516/2020/07/045",
    journal = "JCAP",
    volume = "07",
    number = "07",
    pages = "045",
    year = "2020"
}

@article{Sabogal:2024yha,
    author = "Sabogal, Miguel A. and Silva, Emanuelly and Nunes, Rafael C. and Kumar, Suresh and Di Valentino, Eleonora and Giar{\`e}, William",
    title = "{Quantifying the S8 tension and evidence for interacting dark energy from redshift-space distortion measurements}",
    eprint = "2408.12403",
    archivePrefix = "arXiv",
    primaryClass = "astro-ph.CO",
    doi = "10.1103/PhysRevD.110.123508",
    journal = "Phys. Rev. D",
    volume = "110",
    number = "12",
    pages = "123508",
    year = "2024"
}

@article{Giare:2024smz,
    author = "Giar{\`e}, William and Sabogal, Miguel A. and Nunes, Rafael C. and Di Valentino, Eleonora",
    title = "{Interacting Dark Energy after DESI Baryon Acoustic Oscillation Measurements}",
    eprint = "2404.15232",
    archivePrefix = "arXiv",
    primaryClass = "astro-ph.CO",
    doi = "10.1103/PhysRevLett.133.251003",
    journal = "Phys. Rev. Lett.",
    volume = "133",
    number = "25",
    pages = "251003",
    year = "2024"
}

@article{CosmoVerseNetwork:2025alb,
    author = "Di Valentino, Eleonora and others",
    collaboration = "CosmoVerse Network",
    title = "{The CosmoVerse White Paper: Addressing observational tensions in cosmology with systematics and fundamental physics}",
    eprint = "2504.01669",
    archivePrefix = "arXiv",
    primaryClass = "astro-ph.CO",
    doi = "10.1016/j.dark.2025.101965",
    journal = "Phys. Dark Univ.",
    volume = "49",
    pages = "101965",
    year = "2025"
}

@article{Odintsov:2025kyw,
    author = "Odintsov, Sergei D. and Oikonomou, V. K. and Sharov, German S.",
    title = "{Einstein-Gauss-Bonnet cosmology confronted with observations}",
    eprint = "2503.17946",
    archivePrefix = "arXiv",
    primaryClass = "gr-qc",
    doi = "10.1016/j.jheap.2025.100398",
    journal = "JHEAp",
    volume = "47",
    pages = "100398",
    year = "2025"
}

@article{vanderWesthuizen:2025iam,
    author = "van der Westhuizen, Marcel and Figueruelo, David and Thubisi, Rethabile and Sahlu, Shambel and Abebe, Amare and Paliathanasis, Andronikos",
    title = "{Compartmentalization in the dark sector of the universe after DESI DR2 BAO data}",
    eprint = "2505.23306",
    archivePrefix = "arXiv",
    primaryClass = "astro-ph.CO",
    doi = "10.1016/j.dark.2025.102107",
    journal = "Phys. Dark Univ.",
    volume = "50",
    pages = "102107",
    year = "2025"
}

@article{Paliathanasis:2025dcr,
    author = "Paliathanasis, Andronikos",
    title = "{Dark energy within the generalized uncertainty principle in light of DESI DR2}",
    eprint = "2503.20896",
    archivePrefix = "arXiv",
    primaryClass = "astro-ph.CO",
    doi = "10.1088/1475-7516/2025/09/067",
    journal = "JCAP",
    volume = "09",
    pages = "067",
    year = "2025"
}

@article{Odintsov:2025jfq,
    author = "Odintsov, S. D. and Oikonomou, V. K. and Sharov, G. S.",
    title = "{Dynamical dark energy from F(R) gravity models unifying inflation with dark energy: Confronting the latest observational data}",
    eprint = "2506.02245",
    archivePrefix = "arXiv",
    primaryClass = "gr-qc",
    doi = "10.1016/j.jheap.2025.100471",
    journal = "JHEAp",
    volume = "50",
    pages = "100471",
    year = "2026"
}

@article{Kessler:2025kju,
    author = "Kessler, Daniel A. and Escamilla, Luis A. and Pan, Supriya and Di Valentino, Eleonora",
    title = "{One-parameter dynamical dark energy: Hints for oscillations}",
    eprint = "2504.00776",
    archivePrefix = "arXiv",
    primaryClass = "astro-ph.CO",
    doi = "10.1103/1x6w-4tnd",
    journal = "Phys. Rev. D",
    volume = "114",
    number = "6",
    pages = "063501",
    year = "2026"
}

@article{Capozziello:2005ku,
    author = "Capozziello, S. and Cardone, Vincenzo F. and Troisi, A.",
    title = "{Reconciling dark energy models with f(R) theories}",
    eprint = "astro-ph/0501426",
    archivePrefix = "arXiv",
    doi = "10.1103/PhysRevD.71.043503",
    journal = "Phys. Rev. D",
    volume = "71",
    pages = "043503",
    year = "2005"
}

@article{Hwang:2001pu,
    author = "Hwang, Jai-chan and Noh, Hyerim",
    title = "{f(R) gravity theory and CMBR constraints}",
    eprint = "astro-ph/0102423",
    archivePrefix = "arXiv",
    doi = "10.1016/S0370-2693(01)00404-X",
    journal = "Phys. Lett. B",
    volume = "506",
    pages = "13--19",
    year = "2001"
}

@article{Song:2006ej,
    author = "Song, Yong-Seon and Hu, Wayne and Sawicki, Ignacy",
    title = "{The Large Scale Structure of f(R) Gravity}",
    eprint = "astro-ph/0610532",
    archivePrefix = "arXiv",
    doi = "10.1103/PhysRevD.75.044004",
    journal = "Phys. Rev. D",
    volume = "75",
    pages = "044004",
    year = "2007"
}

@article{Faulkner:2006ub,
    author = "Faulkner, Thomas and Tegmark, Max and Bunn, Emory F. and Mao, Yi",
    title = "{Constraining f(R) Gravity as a Scalar Tensor Theory}",
    eprint = "astro-ph/0612569",
    archivePrefix = "arXiv",
    doi = "10.1103/PhysRevD.76.063505",
    journal = "Phys. Rev. D",
    volume = "76",
    pages = "063505",
    year = "2007"
}

@article{Olmo:2006eh,
    author = "Olmo, Gonzalo J.",
    title = "{Limit to general relativity in f(R) theories of gravity}",
    eprint = "gr-qc/0612047",
    archivePrefix = "arXiv",
    doi = "10.1103/PhysRevD.75.023511",
    journal = "Phys. Rev. D",
    volume = "75",
    pages = "023511",
    year = "2007"
}

@article{Sawicki:2007tf,
    author = "Sawicki, Ignacy and Hu, Wayne",
    title = "{Stability of Cosmological Solution in f(R) Models of Gravity}",
    eprint = "astro-ph/0702278",
    archivePrefix = "arXiv",
    doi = "10.1103/PhysRevD.75.127502",
    journal = "Phys. Rev. D",
    volume = "75",
    pages = "127502",
    year = "2007"
}

@article{Faraoni:2007yn,
    author = "Faraoni, Valerio",
    title = "{de Sitter space and the equivalence between f(R) and scalar-tensor gravity}",
    eprint = "gr-qc/0703044",
    archivePrefix = "arXiv",
    doi = "10.1103/PhysRevD.75.067302",
    journal = "Phys. Rev. D",
    volume = "75",
    pages = "067302",
    year = "2007"
}

@article{Carloni:2007yv,
    author = "Carloni, S. and Dunsby, P. K. S. and Troisi, A.",
    title = "{The Evolution of density perturbations in f(R) gravity}",
    eprint = "0707.0106",
    archivePrefix = "arXiv",
    primaryClass = "gr-qc",
    doi = "10.1103/PhysRevD.77.024024",
    journal = "Phys. Rev. D",
    volume = "77",
    pages = "024024",
    year = "2008"
}

@article{Nojiri:2007as,
    author = "Nojiri, Shin'ichi and Odintsov, Sergei D.",
    title = "{Unifying inflation with LambdaCDM epoch in modified f(R) gravity consistent with Solar System tests}",
    eprint = "0707.1941",
    archivePrefix = "arXiv",
    primaryClass = "hep-th",
    doi = "10.1016/j.physletb.2007.10.027",
    journal = "Phys. Lett. B",
    volume = "657",
    pages = "238--245",
    year = "2007"
}

@article{Deruelle:2007pt,
    author = "Deruelle, Nathalie and Sasaki, Misao and Sendouda, Yuuiti",
    title = "{Junction conditions in f(R) theories of gravity}",
    eprint = "0711.1150",
    archivePrefix = "arXiv",
    primaryClass = "gr-qc",
    reportNumber = "YITP-07-76",
    doi = "10.1143/PTP.119.237",
    journal = "Prog. Theor. Phys.",
    volume = "119",
    pages = "237--251",
    year = "2008"
}

@article{Appleby:2008tv,
    author = "Appleby, S. A. and Battye, R. A.",
    title = "{Aspects of cosmological expansion in F(R) gravity models}",
    eprint = "0803.1081",
    archivePrefix = "arXiv",
    primaryClass = "astro-ph",
    doi = "10.1088/1475-7516/2008/05/019",
    journal = "JCAP",
    volume = "05",
    pages = "019",
    year = "2008"
}

@article{Dunsby:2010wg,
    author = "Dunsby, Peter K. S. and Elizalde, Emilo and Goswami, Rituparno and Odintsov, Sergei and Gomez, Diego Saez",
    title = "{On the LCDM Universe in f(R) gravity}",
    eprint = "1005.2205",
    archivePrefix = "arXiv",
    primaryClass = "gr-qc",
    doi = "10.1103/PhysRevD.82.023519",
    journal = "Phys. Rev. D",
    volume = "82",
    pages = "023519",
    year = "2010"
}

@article{Astashenok:2012tv,
    author = "Astashenok, Artyom V. and Nojiri, Shinichi and Odintsov, Sergei D. and Yurov, Artyom V.",
    title = "{Phantom Cosmology without Big Rip Singularity}",
    eprint = "1201.4056",
    archivePrefix = "arXiv",
    primaryClass = "gr-qc",
    doi = "10.1016/j.physletb.2012.02.039",
    journal = "Phys. Lett. B",
    volume = "709",
    pages = "396--403",
    year = "2012"
}

@article{Odintsov:2026fyc,
    author = "Odintsov, S. D. and Oikonomou, V. K. and Sharov, G. S.",
    title = "{Viable F(R) scenarios unifying inflation with realistic dynamical dark energy}",
    eprint = "2601.06949",
    archivePrefix = "arXiv",
    primaryClass = "gr-qc",
    doi = "10.1016/j.jheap.2026.100579",
    journal = "JHEAp",
    volume = "52",
    pages = "100579",
    year = "2026"
}

@article{Caldwell:1999ew,
    author = "Caldwell, R. R.",
    title = "{A Phantom menace?}",
    eprint = "astro-ph/9908168",
    archivePrefix = "arXiv",
    doi = "10.1016/S0370-2693(02)02589-3",
    journal = "Phys. Lett. B",
    volume = "545",
    pages = "23--29",
    year = "2002"
}

@article{Caldwell:2003vq,
    author = "Caldwell, Robert R. and Kamionkowski, Marc and Weinberg, Nevin N.",
    title = "{Phantom energy and cosmic doomsday}",
    eprint = "astro-ph/0302506",
    archivePrefix = "arXiv",
    doi = "10.1103/PhysRevLett.91.071301",
    journal = "Phys. Rev. Lett.",
    volume = "91",
    pages = "071301",
    year = "2003"
}

@article{Vikman:2004dc,
    author = "Vikman, Alexander",
    title = "{Can dark energy evolve to the phantom?}",
    eprint = "astro-ph/0407107",
    archivePrefix = "arXiv",
    doi = "10.1103/PhysRevD.71.023515",
    journal = "Phys. Rev. D",
    volume = "71",
    pages = "023515",
    year = "2005"
}

@article{Hu:2004kh,
    author = "Hu, Wayne",
    title = "{Crossing the phantom divide: Dark energy internal degrees of freedom}",
    eprint = "astro-ph/0410680",
    archivePrefix = "arXiv",
    doi = "10.1103/PhysRevD.71.047301",
    journal = "Phys. Rev. D",
    volume = "71",
    pages = "047301",
    year = "2005"
}

@article{Carroll:2003st,
    author = "Carroll, Sean M. and Hoffman, Mark and Trodden, Mark",
    title = "{Can the dark energy equation-of-state parameter $w$ be less than $−1$?}",
    eprint = "astro-ph/0301273",
    archivePrefix = "arXiv",
    reportNumber = "EFI-2003-01, SU-GP-03-1-1",
    doi = "10.1103/PhysRevD.68.023509",
    journal = "Phys. Rev. D",
    volume = "68",
    pages = "023509",
    year = "2003"
}

@article{Cline:2003gs,
    author = "Cline, James M. and Jeon, Sangyong and Moore, Guy D.",
    title = "{The Phantom menaced: Constraints on low-energy effective ghosts}",
    eprint = "hep-ph/0311312",
    archivePrefix = "arXiv",
    reportNumber = "MCGILL-03-25",
    doi = "10.1103/PhysRevD.70.043543",
    journal = "Phys. Rev. D",
    volume = "70",
    pages = "043543",
    year = "2004"
}

@article{Nojiri:2005sx,
    author = "Nojiri, Shin'ichi and Odintsov, Sergei D. and Tsujikawa, Shinji",
    title = "{Properties of singularities in (phantom) dark energy universe}",
    eprint = "hep-th/0501025",
    archivePrefix = "arXiv",
    doi = "10.1103/PhysRevD.71.063004",
    journal = "Phys. Rev. D",
    volume = "71",
    pages = "063004",
    year = "2005"
}

@article{Feng:2004ad,
    author = "Feng, Bo and Wang, Xiu-Lian and Zhang, Xin-Min",
    title = "{Dark energy constraints from the cosmic age and supernova}",
    eprint = "astro-ph/0404224",
    archivePrefix = "arXiv",
    doi = "10.1016/j.physletb.2004.12.071",
    journal = "Phys. Lett. B",
    volume = "607",
    pages = "35--41",
    year = "2005"
}

@article{Nojiri:2005sr,
    author = "Nojiri, Shin'ichi and Odintsov, Sergei D.",
    title = "{Inhomogeneous equation of state of the universe: Phantom era, future singularity and crossing the phantom barrier}",
    eprint = "hep-th/0505215",
    archivePrefix = "arXiv",
    doi = "10.1103/PhysRevD.72.023003",
    journal = "Phys. Rev. D",
    volume = "72",
    pages = "023003",
    year = "2005"
}

@article{Kunz:2006wc,
    author = "Kunz, Martin and Sapone, Domenico",
    title = "{Crossing the Phantom Divide}",
    eprint = "astro-ph/0609040",
    archivePrefix = "arXiv",
    doi = "10.1103/PhysRevD.74.123503",
    journal = "Phys. Rev. D",
    volume = "74",
    pages = "123503",
    year = "2006"
}

@article{Bamba:2008hq,
    author = "Bamba, Kazuharu and Geng, Chao-Qiang and Nojiri, Shin'ichi and Odintsov, Sergei D.",
    title = "{Crossing of the phantom divide in modified gravity}",
    eprint = "0810.4296",
    archivePrefix = "arXiv",
    primaryClass = "hep-th",
    doi = "10.1103/PhysRevD.79.083014",
    journal = "Phys. Rev. D",
    volume = "79",
    pages = "083014",
    year = "2009"
}

@article{Cai:2009zp,
    author = "Cai, Yi-Fu and Saridakis, Emmanuel N. and Setare, Mohammad R. and Xia, Jun-Qing",
    title = "{Quintom Cosmology: Theoretical implications and observations}",
    eprint = "0909.2776",
    archivePrefix = "arXiv",
    primaryClass = "hep-th",
    doi = "10.1016/j.physrep.2010.04.001",
    journal = "Phys. Rept.",
    volume = "493",
    pages = "1--60",
    year = "2010"
}

@article{Nojiri:2003vn,
    author = "Nojiri, Shin'ichi and Odintsov, Sergei D.",
    title = "{Quantum de Sitter cosmology and phantom matter}",
    eprint = "hep-th/0303117",
    archivePrefix = "arXiv",
    doi = "10.1016/S0370-2693(03)00594-X",
    journal = "Phys. Lett. B",
    volume = "562",
    pages = "147--152",
    year = "2003"
}

@article{Nojiri:2003jn,
    author = "Nojiri, Shin'ichi and Odintsov, Sergei D.",
    title = "{DeSitter brane universe induced by phantom and quantum effects}",
    eprint = "hep-th/0304131",
    archivePrefix = "arXiv",
    doi = "10.1016/S0370-2693(03)00753-6",
    journal = "Phys. Lett. B",
    volume = "565",
    pages = "1--9",
    year = "2003"
}

@article{Nojiri:2003ag,
    author = "Nojiri, Shin'ichi and Odintsov, Sergei D.",
    title = "{Effective equation of state and energy conditions in phantom / tachyon inflationary cosmology perturbed by quantum effects}",
    eprint = "hep-th/0306212",
    archivePrefix = "arXiv",
    doi = "10.1016/j.physletb.2003.08.013",
    journal = "Phys. Lett. B",
    volume = "571",
    pages = "1--10",
    year = "2003"
}

@article{Elizalde:2004mq,
    author = "Elizalde, Emilio and Nojiri, Shin'ichi and Odintsov, Sergei D.",
    title = "{Late-time cosmology in (phantom) scalar-tensor theory: Dark energy and the cosmic speed-up}",
    eprint = "hep-th/0405034",
    archivePrefix = "arXiv",
    doi = "10.1103/PhysRevD.70.043539",
    journal = "Phys. Rev. D",
    volume = "70",
    pages = "043539",
    year = "2004"
}

@article{Elizalde:2005ju,
    author = "Elizalde, Emilio and Nojiri, Shin'ichi and Odintsov, Sergei D. and Wang, Peng",
    title = "{Dark energy: Vacuum fluctuations, the effective phantom phase, and holography}",
    eprint = "hep-th/0502082",
    archivePrefix = "arXiv",
    doi = "10.1103/PhysRevD.71.103504",
    journal = "Phys. Rev. D",
    volume = "71",
    pages = "103504",
    year = "2005"
}

@article{Nojiri:2005am,
    author = "Nojiri, Shin'ichi and Odintsov, Sergei D. and Gorbunova, O. G.",
    title = "{Dark energy problem: From phantom theory to modified Gauss-Bonnet gravity}",
    eprint = "hep-th/0510183",
    archivePrefix = "arXiv",
    doi = "10.1088/0305-4470/39/21/S62",
    journal = "J. Phys. A",
    volume = "39",
    pages = "6627--6634",
    year = "2006"
}

@article{Briscese:2006xu,
    author = "Briscese, F. and Elizalde, E. and Nojiri, S. and Odintsov, S. D.",
    title = "{Phantom scalar dark energy as modified gravity: Understanding the origin of the Big Rip singularity}",
    eprint = "hep-th/0612220",
    archivePrefix = "arXiv",
    doi = "10.1016/j.physletb.2007.01.013",
    journal = "Phys. Lett. B",
    volume = "646",
    pages = "105--111",
    year = "2007"
}

@article{Elizalde:2008yf,
    author = "Elizalde, Emilio and Nojiri, Shin'ichi and Odintsov, Sergei D. and Saez-Gomez, Diego and Faraoni, Valerio",
    title = "{Reconstructing the universe history, from inflation to acceleration, with phantom and canonical scalar fields}",
    eprint = "0803.1311",
    archivePrefix = "arXiv",
    primaryClass = "hep-th",
    doi = "10.1103/PhysRevD.77.106005",
    journal = "Phys. Rev. D",
    volume = "77",
    pages = "106005",
    year = "2008"
}

@article{Bamba:2010bm,
    author = "Bamba, Kazuharu and Geng, Chao-Qiang and Nojiri, Shin'ichi and Odintsov, Sergei D.",
    editor = "Bamba, Kazuharu and Geng, Chao-Qiang",
    title = "{Crossing of Phantom Divide in $F(R)$ Gravity}",
    eprint = "1003.0769",
    archivePrefix = "arXiv",
    primaryClass = "hep-th",
    doi = "10.1142/S0217732310000058",
    journal = "Mod. Phys. Lett. A",
    volume = "25",
    pages = "900--908",
    year = "2010"
}

@article{Bamba:2012ky,
    author = "Bamba, Kazuharu and Nojiri, Shinichi and Odintsov, Sergei D. and Sasaki, Misao",
    title = "{Screening of cosmological constant for De Sitter Universe in non-local gravity, phantom-divide crossing and finite-time future singularities}",
    eprint = "1104.2692",
    archivePrefix = "arXiv",
    primaryClass = "hep-th",
    reportNumber = "YITP-11-46",
    doi = "10.1007/s10714-012-1342-7",
    journal = "Gen. Rel. Grav.",
    volume = "44",
    pages = "1321--1356",
    year = "2012"
}

@article{Saitou:2012xw,
    author = "Saitou, Rio and Nojiri, Shin'ichi",
    title = "{Stable phantom-divide crossing in two scalar models with matter}",
    eprint = "1203.1442",
    archivePrefix = "arXiv",
    primaryClass = "hep-th",
    doi = "10.1140/epjc/s10052-012-1946-9",
    journal = "Eur. Phys. J. C",
    volume = "72",
    pages = "1946",
    year = "2012"
}

@article{Nojiri:2013ru,
    author = "Nojiri, Shin'ichi and Saridakis, Emmanuel N.",
    title = "{Phantom without ghost}",
    eprint = "1301.2686",
    archivePrefix = "arXiv",
    primaryClass = "hep-th",
    doi = "10.1007/s10509-013-1509-z",
    journal = "Astrophys. Space Sci.",
    volume = "347",
    pages = "221--226",
    year = "2013"
}

@article{Nojiri:2015fia,
    author = "Nojiri, Shin'ichi and Odintsov, S. D. and Oikonomou, V. K. and Saridakis, Emmanuel N.",
    title = "{Singular cosmological evolution using canonical and ghost scalar fields}",
    eprint = "1503.08443",
    archivePrefix = "arXiv",
    primaryClass = "gr-qc",
    doi = "10.1088/1475-7516/2015/9/044",
    journal = "JCAP",
    volume = "09",
    pages = "044",
    year = "2015"
}

@article{Cardenas:2020grl,
    author = "C{\'a}rdenas, V{\'\i}ctor H. and Cruz, Miguel and Lepe, Samuel and Nojiri, Shin'ichi and Odintsov, Sergei D.",
    title = "{Challenging matter creation models in the phantom divide}",
    eprint = "2004.02337",
    archivePrefix = "arXiv",
    primaryClass = "gr-qc",
    doi = "10.1103/PhysRevD.101.083530",
    journal = "Phys. Rev. D",
    volume = "101",
    number = "8",
    pages = "083530",
    year = "2020"
}

@article{Odintsov:2024lid,
    author = "Odintsov, Sergei D. and S{\'a}ez-Chill{\'o}n G{\'o}mez, Diego and Sharov, German S.",
    title = "{Exponential gravity with logarithmic corrections in the presence of axion dark matter}",
    eprint = "2406.08831",
    archivePrefix = "arXiv",
    primaryClass = "gr-qc",
    doi = "10.1016/j.dark.2024.101558",
    journal = "Phys. Dark Univ.",
    volume = "46",
    pages = "101558",
    year = "2024"
}

@article{Capozziello:2002rd,
    author = "Capozziello, Salvatore",
    title = "{Curvature quintessence}",
    eprint = "gr-qc/0201033",
    archivePrefix = "arXiv",
    doi = "10.1142/S0218271802002025",
    journal = "Int. J. Mod. Phys. D",
    volume = "11",
    pages = "483--492",
    year = "2002"
}

@article{Cognola:2007zu,
    author = "Cognola, G. and Elizalde, E. and Nojiri, S. and Odintsov, S. D. and Sebastiani, L. and Zerbini, S.",
    title = "{A Class of viable modified f(R) gravities describing inflation and the onset of accelerated expansion}",
    eprint = "0712.4017",
    archivePrefix = "arXiv",
    primaryClass = "hep-th",
    doi = "10.1103/PhysRevD.77.046009",
    journal = "Phys. Rev. D",
    volume = "77",
    pages = "046009",
    year = "2008"
}

@article{Appleby:2009uf,
    author = "Appleby, Stephen A. and Battye, Richard A. and Starobinsky, Alexei A.",
    title = "{Curing singularities in cosmological evolution of F(R) gravity}",
    eprint = "0909.1737",
    archivePrefix = "arXiv",
    primaryClass = "astro-ph.CO",
    doi = "10.1088/1475-7516/2010/06/005",
    journal = "JCAP",
    volume = "06",
    pages = "005",
    year = "2010"
}

@article{Elizalde:2010ts,
    author = "Elizalde, E. and Nojiri, S. and Odintsov, S. D. and Sebastiani, L. and Zerbini, S.",
    title = "{Non-singular exponential gravity: a simple theory for early- and late-time accelerated expansion}",
    eprint = "1012.2280",
    archivePrefix = "arXiv",
    primaryClass = "hep-th",
    doi = "10.1103/PhysRevD.83.086006",
    journal = "Phys. Rev. D",
    volume = "83",
    pages = "086006",
    year = "2011"
}

@article{Odintsov:2017hbk,
    author = "Odintsov, S. D. and Oikonomou, V. K. and Sebastiani, L.",
    title = "{Unification of Constant-roll Inflation and Dark Energy with Logarithmic $R^2$-corrected and Exponential $F(R)$ Gravity}",
    eprint = "1708.08346",
    archivePrefix = "arXiv",
    primaryClass = "gr-qc",
    doi = "10.1016/j.nuclphysb.2017.08.018",
    journal = "Nucl. Phys. B",
    volume = "923",
    pages = "608--632",
    year = "2017"
}

@article{Nojiri:2019fft,
    author = "Nojiri, Shin'ichi and Odintsov, Sergei D. and Oikonomou, V. K.",
    title = "{Unifying Inflation with Early and Late-time Dark Energy in $F(R)$ Gravity}",
    eprint = "1912.13128",
    archivePrefix = "arXiv",
    primaryClass = "gr-qc",
    doi = "10.1016/j.dark.2020.100602",
    journal = "Phys. Dark Univ.",
    volume = "29",
    pages = "100602",
    year = "2020"
}

@article{Frampton:2011sp,
    author = "Frampton, Paul H. and Ludwick, Kevin J. and Scherrer, Robert J.",
    title = "{The Little Rip}",
    eprint = "1106.4996",
    archivePrefix = "arXiv",
    primaryClass = "astro-ph.CO",
    doi = "10.1103/PhysRevD.84.063003",
    journal = "Phys. Rev. D",
    volume = "84",
    pages = "063003",
    year = "2011"
}

@article{Brevik:2011mm,
    author = "Brevik, I. and Elizalde, E. and Nojiri, S. and Odintsov, S. D.",
    title = "{Viscous Little Rip Cosmology}",
    eprint = "1107.4642",
    archivePrefix = "arXiv",
    primaryClass = "hep-th",
    doi = "10.1103/PhysRevD.84.103508",
    journal = "Phys. Rev. D",
    volume = "84",
    pages = "103508",
    year = "2011"
}

@article{Nojiri:2011kd,
    author = "Nojiri, Shinichi and Odintsov, S. D. and Saez-Gomez, D.",
    editor = "Beltran Jimenez, Jose and Ruiz Cembranos, Jose Alberto and Dobado, Antonio and Lopez Maroto, Antonio and De la Cruz Dombriz, A.",
    title = "{Cyclic, ekpyrotic and little rip universe in modified gravity}",
    eprint = "1108.0767",
    archivePrefix = "arXiv",
    primaryClass = "hep-th",
    reportNumber = "AEI-2011-051",
    doi = "10.1063/1.4734414",
    journal = "AIP Conf. Proc.",
    volume = "1458",
    number = "1",
    pages = "207--221",
    year = "2012"
}

@article{Frampton:2011rh,
    author = "Frampton, Paul H. and Ludwick, Kevin J. and Nojiri, Shin'ichi and Odintsov, Sergei D. and Scherrer, Robert J.",
    title = "{Models for Little Rip Dark Energy}",
    eprint = "1108.0067",
    archivePrefix = "arXiv",
    primaryClass = "hep-th",
    doi = "10.1016/j.physletb.2012.01.048",
    journal = "Phys. Lett. B",
    volume = "708",
    pages = "204--211",
    year = "2012"
}

@article{Frampton:2011aa,
    author = "Frampton, Paul H. and Ludwick, Kevin J. and Scherrer, Robert J.",
    title = "{Pseudo-rip: Cosmological models intermediate between the cosmological constant and the little rip}",
    eprint = "1112.2964",
    archivePrefix = "arXiv",
    primaryClass = "astro-ph.CO",
    doi = "10.1103/PhysRevD.85.083001",
    journal = "Phys. Rev. D",
    volume = "85",
    pages = "083001",
    year = "2012"
}

@article{Bamba:2011sm,
    author = "Bamba, Kazuharu and Nojiri, Shin'ichi and Odintsov, Sergei D.",
    title = "{Time-dependent matter instability and star singularity in $F(R)$ gravity}",
    eprint = "1101.2820",
    archivePrefix = "arXiv",
    primaryClass = "gr-qc",
    doi = "10.1016/j.physletb.2011.03.038",
    journal = "Phys. Lett. B",
    volume = "698",
    pages = "451--456",
    year = "2011"
}

@article{Barrow:2004xh,
    author = "Barrow, John D.",
    title = "{Sudden future singularities}",
    eprint = "gr-qc/0403084",
    archivePrefix = "arXiv",
    doi = "10.1088/0264-9381/21/11/L03",
    journal = "Class. Quant. Grav.",
    volume = "21",
    pages = "L79--L82",
    year = "2004"
}

@article{Nojiri:2004pf,
    author = "Nojiri, Shin'ichi and Odintsov, Sergei D.",
    title = "{The Final state and thermodynamics of dark energy universe}",
    eprint = "hep-th/0408170",
    archivePrefix = "arXiv",
    doi = "10.1103/PhysRevD.70.103522",
    journal = "Phys. Rev. D",
    volume = "70",
    pages = "103522",
    year = "2004"
}

@article{deHaro:2023lbq,
    author = "de Haro, Jaume and Nojiri, Shin'ichi and Odintsov, S. D. and Oikonomou, V. K. and Pan, Supriya",
    title = "{Finite-time cosmological singularities and the possible fate of the Universe}",
    eprint = "2309.07465",
    archivePrefix = "arXiv",
    primaryClass = "gr-qc",
    doi = "10.1016/j.physrep.2023.09.003",
    journal = "Phys. Rept.",
    volume = "1034",
    pages = "1--114",
    year = "2023"
}

@article{Nojiri:2026uvn,
    author = "Nojiri, Shin'ichi and Odintsov, S. D. and Oikonomou, V. K.",
    title = "{Is Phantom Divide Crossing in General Relativity Completely Impossible? Shortcomings in Canonical and Minimally Coupled Scalar Field and Possible Solutions ink-Essence Models}",
    eprint = "2601.21356",
    archivePrefix = "arXiv",
    primaryClass = "gr-qc",
    reportNumber = "KEK-TH-2796, KEK-Cosmo-0406",
    doi = "10.1002/prop.70137",
    journal = "Fortsch. Phys.",
    volume = "74",
    number = "8",
    pages = "e70137",
    year = "2026"
}

@article{Nojiri:2025uew,
    author = "Nojiri, Shin'ichi and Odintsov, Sergei D. and Oikonomou, V. K.",
    title = "{Apparent phantom crossing in Gauss{\textendash}Bonnet gravity}",
    eprint = "2512.06279",
    archivePrefix = "arXiv",
    primaryClass = "gr-qc",
    reportNumber = "KEK-TH-2786, KEK-Cosmo-0404",
    doi = "10.1140/epjc/s10052-026-15562-x",
    journal = "Eur. Phys. J. C",
    volume = "86",
    number = "4",
    pages = "353",
    year = "2026"
}

@article{Kiefer:2008sw,
    author = "Kiefer, Claus and Sandhoefer, Barbara",
    title = "{Quantum cosmology}",
    eprint = "0804.0672",
    archivePrefix = "arXiv",
    primaryClass = "gr-qc",
    reportNumber = "Cologne preprint GR-08/2, COLOGNE-PREPRINT-GR-08-2",
    doi = "10.1515/zna-2021-0384",
    journal = "Z. Naturforsch. A",
    volume = "77",
    number = "6",
    pages = "543--559",
    year = "2022"
}

@article{Kamenshchik:2007zj,
    author = "Kamenshchik, Alexander and Kiefer, Claus and Sandhofer, Barbara",
    title = "{Quantum cosmology with big-brake singularity}",
    eprint = "0705.1688",
    archivePrefix = "arXiv",
    primaryClass = "gr-qc",
    doi = "10.1103/PhysRevD.76.064032",
    journal = "Phys. Rev. D",
    volume = "76",
    pages = "064032",
    year = "2007"
}

@article{Bouhmadi-Lopez:2009ggt,
    author = "Bouhmadi-Lopez, Mariam and Kiefer, Claus and Sandhofer, Barbara and Vargas Moniz, Paulo",
    title = "{On the quantum fate of singularities in a dark-energy dominated universe}",
    eprint = "0905.2421",
    archivePrefix = "arXiv",
    primaryClass = "gr-qc",
    doi = "10.1103/PhysRevD.79.124035",
    journal = "Phys. Rev. D",
    volume = "79",
    pages = "124035",
    year = "2009"
}

@article{Alonso-Serrano:2018zpi,
    author = "Alonso-Serrano, Ana and Bouhmadi-L{\'o}pez, Mariam and Mart{\'\i}n-Moruno, Prado",
    title = "{$f(R)$ quantum cosmology: avoiding the Big Rip}",
    eprint = "1802.03290",
    archivePrefix = "arXiv",
    primaryClass = "gr-qc",
    doi = "10.1103/PhysRevD.98.104004",
    journal = "Phys. Rev. D",
    volume = "98",
    number = "10",
    pages = "104004",
    year = "2018"
}

@article{BorislavovVasilev:2021elf,
    author = "Borislavov Vasilev, Teodor and Bouhmadi-L{\'o}pez, Mariam and Mart{\'\i}n-Moruno, Prado",
    title = "{Little rip in classical and quantum f(R) cosmology}",
    eprint = "2103.12786",
    archivePrefix = "arXiv",
    primaryClass = "gr-qc",
    doi = "10.1103/PhysRevD.103.124049",
    journal = "Phys. Rev. D",
    volume = "103",
    number = "12",
    pages = "124049",
    year = "2021"
}

@article{Kiefer:2019bxk,
    author = "Kiefer, Claus and Kwidzinski, Nick and Piontek, Dennis",
    title = "{Singularity avoidance in Bianchi I quantum cosmology}",
    eprint = "1903.04391",
    archivePrefix = "arXiv",
    primaryClass = "gr-qc",
    doi = "10.1140/epjc/s10052-019-7193-6",
    journal = "Eur. Phys. J. C",
    volume = "79",
    number = "8",
    pages = "686",
    year = "2019"
}

@article{Hu:2007nk,
    author = "Hu, Wayne and Sawicki, Ignacy",
    title = "{Models of f(R) Cosmic Acceleration that Evade Solar-System Tests}",
    eprint = "0705.1158",
    archivePrefix = "arXiv",
    primaryClass = "astro-ph",
    doi = "10.1103/PhysRevD.76.064004",
    journal = "Phys. Rev. D",
    volume = "76",
    pages = "064004",
    year = "2007"
}

@article{Koyama:2015vza,
    author = "Koyama, Kazuya",
    title = "{Cosmological Tests of Modified Gravity}",
    eprint = "1504.04623",
    archivePrefix = "arXiv",
    primaryClass = "astro-ph.CO",
    doi = "10.1088/0034-4885/79/4/046902",
    journal = "Rept. Prog. Phys.",
    volume = "79",
    number = "4",
    pages = "046902",
    year = "2016"
}

@article{Mueller:2014nsa,
    author = "Mueller, Eva-Maria and de Bernardis, Francesco and Bean, Rachel and Niemack, Michael D.",
    title = "{Constraints on gravity and dark energy from the pairwise kinematic Sunyaev-Zeldovich effect}",
    eprint = "1408.6248",
    archivePrefix = "arXiv",
    primaryClass = "astro-ph.CO",
    doi = "10.1088/0004-637X/808/1/47",
    journal = "Astrophys. J.",
    volume = "808",
    number = "1",
    pages = "47",
    year = "2015"
}

@article{Bianchini:2015iaa,
    author = "Bianchini, Federico and Silvestri, Alessandra",
    title = "{Kinetic Sunyaev-Zel{\textquoteright}dovich effect in modified gravity}",
    eprint = "1510.08844",
    archivePrefix = "arXiv",
    primaryClass = "astro-ph.CO",
    doi = "10.1103/PhysRevD.93.064026",
    journal = "Phys. Rev. D",
    volume = "93",
    number = "6",
    pages = "064026",
    year = "2016"
}

@article{Hand:2012ui,
    author = "Hand, Nick and others",
    title = "{Evidence of Galaxy Cluster Motions with the Kinematic Sunyaev-Zel'dovich Effect}",
    eprint = "1203.4219",
    archivePrefix = "arXiv",
    primaryClass = "astro-ph.CO",
    doi = "10.1103/PhysRevLett.109.041101",
    journal = "Phys. Rev. Lett.",
    volume = "109",
    pages = "041101",
    year = "2012"
}

@article{Flender:2015btu,
    author = "Flender, Samuel and Bleem, Lindsey and Finkel, Hal and Habib, Salman and Heitmann, Katrin and Holder, Gilbert",
    title = "{Simulations of the Pairwise Kinematic Sunyaev{\textendash}Zel{\textquoteright}dovich Signal}",
    eprint = "1511.02843",
    archivePrefix = "arXiv",
    primaryClass = "astro-ph.CO",
    doi = "10.3847/0004-637X/823/2/98",
    journal = "Astrophys. J.",
    volume = "823",
    number = "2",
    pages = "98",
    year = "2016"
}

@article{Gong:2025ffw,
    author = "Gong, Yulin and others",
    title = "{Detection of the pairwise kinematic Sunyaev-Zel{\textquoteright}dovich effect and pairwise velocity with DESI DR1 galaxies and ACT DR6 and Planck CMB data}",
    eprint = "2511.23417",
    archivePrefix = "arXiv",
    primaryClass = "astro-ph.CO",
    reportNumber = "FERMILAB-PUB-25-0868-PPD",
    doi = "10.1103/kkrh-1svk",
    journal = "Phys. Rev. D",
    volume = "113",
    number = "6",
    pages = "063538",
    year = "2026"
}

@article{Lai:2025xkf,
    author = "Lai, Y. and others",
    title = "{The DESI DR1 peculiar velocity survey: growth rate measurements from the maximum likelihood fields method}",
    eprint = "2512.03229",
    archivePrefix = "arXiv",
    primaryClass = "astro-ph.CO",
    reportNumber = "FERMILAB-PUB-25-0900-PPD",
    doi = "10.1088/1475-7516/2026/04/026",
    journal = "JCAP",
    volume = "04",
    pages = "026",
    year = "2026"
}

@article{Hadzhiyska:2025egz,
    author = "Hadzhiyska, B. and others",
    title = "{Probing cosmic velocities with the pairwise kinematic Sunyaev-Zel{\textquoteright}dovich signal in DESI Bright Galaxy Sample DR1 and ACT DR6}",
    eprint = "2510.14135",
    archivePrefix = "arXiv",
    primaryClass = "astro-ph.CO",
    reportNumber = "FERMILAB-PUB-25-0758-PPD",
    doi = "10.1103/9h6g-gxdn",
    journal = "Phys. Rev. D",
    volume = "113",
    number = "6",
    pages = "063565",
    year = "2026"
}

@article{Trivedi:2023zlf,
    author = "Trivedi, Oem",
    title = "{Recent Advances in Cosmological Singularities}",
    eprint = "2309.08954",
    archivePrefix = "arXiv",
    primaryClass = "gr-qc",
    doi = "10.3390/sym16030298",
    journal = "Symmetry",
    volume = "16",
    number = "3",
    pages = "298",
    year = "2024"
}

\end{document}